\documentclass[12pt]{elsarticle}                                          
\usepackage{graphicx}                                                         
\usepackage{a41}                                                                
\usepackage{xcolor}                     
\usepackage[rflt]{floatflt}
\usepackage{float}
\usepackage{lscape}
\usepackage{array}
\usepackage[english]{babel}
\usepackage[T1]{fontenc}
\usepackage{ae}
\usepackage{url}
\usepackage{amsmath, amsthm, amssymb}
\usepackage{slashed}

\usepackage{rotating}
\usepackage{graphicx}
\usepackage{comment}
\newcounter{mmacnt}
\def\restartmma{\setcounter{mmacnt}{0}}
\restartmma \catcode`|=\active
\def|#1|{\mathrm{#1}}
\catcode`|=12
\newenvironment{mma}{
\par\smallskip
\catcode`|=\active
\parskip=0pt\parindent=0pt 
\small
\def\In##1\\{%
\def\linebreak{\hfill\break\null\qquad}%
\refstepcounter{mmacnt}
\hangindent=2.5em\hangafter=0
\leavevmode
\llap{\tiny\sffamily In[\arabic{mmacnt}]:=\kern.5em}%
\mathversion{bold}\footnotesize$
\displaystyle##1$\normalsize
\mathversion{normal}\par
 }%
\def\Print##1\\{%
\def\linebreak{\hfill\break}%
\hangindent=2.5em\hangafter=0
\leavevmode ##1\par}%
\def\Out##1\\{%
\def\linebreak{$\hfill\break\null\hfill$}%
\kern\abovedisplayskip\par
\hangindent=2.5em\hangafter=0
\leavevmode
\llap{\tiny\sffamily Out[\arabic{mmacnt}]=\kern.5em}
\footnotesize$\displaystyle##1$
\normalsize\hfill\null\par
\kern\belowdisplayskip
}%
\def\Warning##1##2\\{%
\def\linebreak{\hfill\break}%
\hangindent=2.5em\hangafter=0
\leavevmode
{\scriptsize##1 : ##2}\par}%
}{%
\par\smallskip
}

\usepackage{color}
\newenvironment{fshaded}{%
\MakeFramed {\FrameRestore}
}%
{\endMakeFramed}

\makeatletter
\def\ps@pprintTitle{%
\let\@oddhead\@empty
\let\@evenhead\@empty
\def\@oddfoot{\reset@font\hfil\thepage\hfil}
\let\@evenfoot\@oddfoot
}
\makeatother
\usepackage{tikz}
\usetikzlibrary{matrix}
\allowdisplaybreaks[4]

\begin{document}
\begin{frontmatter}
\title{\Large
\textbf{
Charged Higgs bosons associated 
with neutral gauge bosons at future 
multi--TeV muon colliders
}}
\author[1,2]{Khiem Hong Phan}
\ead{phanhongkhiem@duytan.edu.vn}
\author[5]{Quang Hoang-Minh Pham}
\address[1]{\it Institute of Fundamental
and Applied Sciences, Duy Tan University,
Ho Chi Minh City $70000$, Vietnam}
\address[2]{Faculty of Natural Sciences,
Duy Tan University, Da Nang City $50000$,
Vietnam}
\address[5]
{\it VNUHCM-University of Science,
$227$ Nguyen Van Cu, District $5$,
Ho Chi Minh City $700000$, Vietnam}
\pagestyle{myheadings}
\markright{}
\begin{abstract} 
The first results for charged Higgs pair production 
associated with neutral gauge bosons at 
future multi--TeV muon colliders are presented 
within the framework of the 
Two-Higgs-Doublet Model. In the phenomenological studies 
for the production processes, we first update the parameter 
space of the Type-X Two-Higgs-Doublet Model. From the viable 
regions of the parameter space, we compute the production 
cross sections, including all two-body decay modes of 
the charged Higgs boson. 
We mention that all two-body decay modes of the 
charged Higgs boson are evaluated with one-loop 
electroweak corrections, with two-loop QCD corrections 
also included for the decays of the charged Higgs 
into two fermions. The signal significances are also 
examined for several benchmark points within the
updated parameter space. It is worth emphasizing 
that the production cross sections used for computing the
significances in this work include initial-state radiation
corrections up to two-loop order.
With the help of the high integrated luminosity planned 
for future multi--TeV muon colliders, the production 
signals can be observed with a statistical significance 
exceeding $5\sigma$ for several selected points 
within the allowed parameter space.
\end{abstract}
\begin{keyword} 
\footnotesize
Scalar Higgs bosons
phenomenology, Physics 
beyond the Standard Model,
Physics at present
and future multi--TeV muon colliders. 
\end{keyword}
\end{frontmatter}
\section{Introduction}
Searches for additional scalar particles predicted 
in physics beyond the Standard Model (SM) are among 
the main goals of future colliders, including the 
High-Luminosity Large Hadron 
Collider (HL-LHC), the High-Energy LHC (HE-LHC), 
future lepton colliders such as 
the International Linear Collider, and multi--TeV muon colliders. 
The discovery of additional scalar particles could provide 
a direct signal of new physics and offer important insights 
for verifying the
scalar potential in various particle models. Such data therefore
provide crucial information to deepen our understanding of
electroweak symmetry breaking in high-energy physics.
Among all scalar particle productions, searches for charged Higgs
boson production and its decay channels have recently been widely
studied at current and future colliders. It is worth
mentioning typical searches for charged Higgs bosons in the
following paragraphs. At the LHC, charged Higgs bosons can be
probed through top quark decays, as reported in
\cite{CMS:2012fgz, ATLAS:2023bzb, ATLAS:2024oqu}. 
In $pp$ collisions at $\sqrt{s} = 8$ TeV, charged Higgs
boson production has been measured, as reported in \cite{CMS:2015lsf}.
Measurements of charged Higgs bosons from di-top quark events
using $pp$ collision data at $\sqrt{s} = 7$ TeV have been
performed with considering the sequential decay $H^+ \to \tau\nu$
in \cite{ATLAS:2012nhc, ATLAS:2018gfm, ATLAS:2024hya, CMS:2019bfg},
as well as the decay mode $H^+ \to c\bar{s}$ in
\cite{ATLAS:2013uxj, CMS:2020osd}.
Investigations of the heavy charged Higgs mass regime through the
process $pp \to tH^\pm \to t\bar{t}b$ have been conducted in
\cite{ATLAS:2015nkq, ATLAS:2021upq, CMS:2020imj}. Furthermore,
charged Higgs productions via vector boson fusions, considering the
decay mode $H^\pm \to W^\pm Z$, havae been studied in
\cite{ATLAS:2015edr, CMS:2021wlt}. Measurements of charm and bottom
quarks produced by charged Higgs decays have also been performed
at the LHC \cite{CMS:2018dzl}. In addition, charged Higgs decays
to a heavy CP-even Higgs as well as to the SM-like Higgs boson
at the LHC have been measured, as reported in
\cite{CMS:2022jqc, ATLAS:2024rcu}.

There exist many theoretical computations for charged Higgs production
at current and future colliders. Single charged Higgs production
via the process $pp \to tH^- \to tW^-b\bar{b}$ at the LHC 
has been studied in~\cite{Arhrib:2017veb}.
Investigations of charged Higgs pair production at the HL-LHC
and the HE-LHC, including their subsequent bosonic decay channels,
can be found in~\cite{Arhrib:2019ywg}. Within the MSSM scenarios,
the dominant channel of single charged Higgs production, such as
$gg \to tbH$, has been examined at the LHC, as reported in
\cite{Arhrib:2019ykh}. Light charged Higgs bosons decaying into
bosonic modes have been investigated, including implications from
LHC Run III data within the framework of the THDM, as published
in~\cite{Arhrib:2020tqk}. Furthermore, searches for light charged
Higgs at the LHC through the channels $pp \to H^\pm h/A$ and
$pp \to H^+H^-$, with $H^\pm \to W^\pm h/A$, were performed in
\cite{Arhrib:2021xmc}. Other searches for charged Higgs at the
LHC have been carried out in \cite{Wang:2021pxc,
Krab:2022lih, Arhrib:2024sfg, Arhrib:2024nbj, Logan:2018wtm}.
At future lepton colliders, including $e^-e^+$ and $\mu^-\mu^+$
machines, charged Higgs bosons have been studied in
\cite{Ouazghour:2023plc, Ouazghour:2024twx, Ouazghour:2025owf,
BrahimAit-Ouazghour:2025mhy, Ahmed:2024oxg, Hashemi:2023osd}.
In our previous works, we discussed the possibility of probing
charged Higgs bosons, including $H^\pm \to W^\pm \gamma$ in
\cite{Tran:2025iur} and $H^\pm \to W^\pm Z$ in
\cite{Tran:2025zfq} at future multi--TeV muon colliders.

Future lepton colliders are well known for providing a 
cleaner environment than hadron colliders, which have 
large QCD backgrounds, allowing for higher-precision 
measurements. Moreover,
multi--TeV muon colliders could provide access to a higher-energy
regime, planned for probing new physics. In comparison with
$e^-e^+$ collisions, contributions from scalar particles exchanged
in $s$-channels may be enhanced due to resonance effects, which
highlight the advantage of multi--TeV muon colliders as important
tools for exploring and discovering the scalar Higgs sectors.
In the four types of THDM, type-I and type-X allow
charged Higgs masses in a wide range, from $\mathcal{O}(100)$ GeV to
$\mathcal{O}(1000)$ GeV, in comparison with the other
types, for which charged Higgs masses are constrained above
$500$ GeV. Therefore, both type-I and type-X are more
interesting for searches in the low-Higgs mass regions than
the other types.

In this work, we present the first results
for charged Higgs pair production associated 
with neutral gauge bosons at future multi--TeV 
muon colliders within the framework of the 
THDM. In the phenomenological studies 
for the production processes, we first update the parameter 
space of the Type-X THDM. From the viable 
regions of the parameter space, we compute the production 
cross sections, including all two-body decay modes of 
the charged Higgs boson. 
We emphasize that all two-body decay modes of the charged 
Higgs boson are evaluated at the one-loop level, with up to 
two-loop QCD corrections included for decays of the charged 
Higgs into two fermions.
The signal significances are also 
examined for several benchmark points within the
updated parameter space. It is worth mentioning that the
production cross sections used for computing the
significances in this work include initial-state radiation
corrections up to two-loop order. With the help of 
the high luminosity planned at future multi--TeV 
muon colliders, the signals can be observed with a 
statistical significance exceeding $5\sigma$ for 
several benchmark points within the
viable parameter space.

The layout of our paper is as follows. We briefly present
the model under study, known as the THDM, and focus on
its constraints for Type-X in the next section. Section~3
is devoted to the phenomenological study of charged Higgs
pair production associated with neutral gauge bosons at
future multi--TeV muon colliders. In Section~4, the same
production is investigated via $\gamma\gamma$-fusion 
at future
multi--TeV muon colliders. 
The signal significances are also 
examined for several benchmark points within the
updated parameter space are computed in Section~5.
Finally, the conclusions and outlooks
of our work are presented in Section~6. 
\section{The Two-Higgs-Doublet Model
and Its Constraints}
In the model, the fermion and gauge 
sectors remain the same
as in the SM. The scalar sector is extended by
introducing an additional scalar Higgs field. For the THDM, 
we refer readers to~Ref.~\cite{Branco:2011iw} for detailed 
reviews and phenomenological studies. Moreover, we consider
the scalar potential preserving
the discrete $Z_2$ symmetry, given by $\Phi_1 \leftrightarrow
\Phi_1$ and $\Phi_2 \leftrightarrow -\Phi_2$, with soft-breaking
terms allowed, as seen in~\cite{Aoki:2009ha}. In the present
study, we focus on the THDM with CP-conserving case. As a
result, all parameters in the scalar potential are real variables. 
In the Yukawa sector, due to the $Z_2$ symmetry and
depending on the $Z_2$ charge assignments, we classify the THDM
into four types, known as Type-I, Type-II, Type-X, and Type-Y
~\cite{Aoki:2009ha}. Consequently, the Lagrangian for Yukawa
interactions can be parameterized as follows 
(taken from Refs.~\cite{Tran:2025iur, Tran:2025zfq}):
\begin{eqnarray}
{\mathcal L}_\text{Y}
&=&
-\sum_{f=u,d,\ell}
\left(
\sum_{\phi_j=h, H}
g_{\phi_j ff}\cdot
\phi_j{\overline f}f
+
g_{Aff}\cdot
A
{\overline f}
\gamma_5f
\right)
\\
&&
-
\left[
\bar{u}_{i}
\left(
g_{H^+ u_i d_j}^L m_{u_i}
P_L
+
g_{H^+ u_i d_j}^R
m_{d_j} P_R \right)d_{j} H^+
\right]
\nonumber\\
&&
\nonumber\\
&&
+
\cdots
\nonumber
\\
&=&
-\sum_{f=u,d,\ell}
\left(
\sum_{\phi_j=h, H}
\frac{m_f}{v}\xi_{\phi_j}^f
\phi_j {\overline f}f
-i\frac{m_f}{v}\xi_A^f
{\overline f}
\gamma_5fA
\right)
\\
&&
-
\frac{
\sqrt{2}
}{v}
\left[
\bar{u}_{i}
V_{ij}\left(
m_{u_i}
\xi^{u}_A P_L
+
\xi^{d}_A
m_{d_j} P_R \right)d_{j} H^+
\right]
\nonumber\\
&&
- \frac{\sqrt{2}}{v}
\bar{\nu}_L
\xi^{\ell}_A
m_\ell \ell_R H^+
+ \textrm{H.c}.
\nonumber
\end{eqnarray}
Here, $\ell_{L/R}$ and $\nu_{L/R}$ denote the 
left- and right-handed leptons and neutrinos, 
respectively. The factor $V_{ij}$ represents
an element of the CKM matrix. The projection 
operators are defined as $P_{L/R} = (1 \mp \gamma_{5})/2$, 
which are used in the Lagrangian.
We note that the coupling coefficients appearing 
in the Lagrangian are listed in Table~2 of 
Ref.~\cite{Aoki:2009ha}.

We now present the constraints on the THDM. The models
under investigation are restricted by both theoretical
conditions and experimental data from collider
measurements. These aspects are discussed in more detail
in our previous works~\cite{Tran:2025iur, Tran:2025zfq}.
In this study, we focus on the phenomenological analysis
of the Type-X scenario. Accordingly, we only show the
allowed parameter space for this type of THDM. We first
choose the parameters for the Type-X THDM within the
ranges $s_{\beta-\alpha} \in [0.97, 1]$, $t_{\beta} \in
[0.5, 45]$, $m_{H} \in [130, 1000]~\text{GeV}$,
$m_{A,H^{\pm}} \in [130, 1000]~\text{GeV}$, and
$m_{12}^2 \in [0, 10^6]~\text{GeV}^2$, with the SM-like
Higgs mass fixed at $m_{h} = 125.09~\text{GeV}$. The
scanning of the parameters above is performed as follows.
We first constrain the parameter space using theoretical
requirements, including unitarity, perturbativity, and
vacuum stability of the scalar potential. The surviving
points are then tested against the electroweak precision
observables (EWPOs). All remaining points are subsequently
compared with data on the SM-like Higgs boson properties
and the exclusion limits for BSM Higgs bosons measured at
the LHC. For all the aforementioned constraints, we use
{\tt 2HDMC}--1.8.0~\cite{Eriksson:2009ws} to scan the
parameter space. Finally, the surviving parameter points
are further constrained by flavor data, which are taken
into account using the {\tt SuperISO}
package~\cite{Mahmoudi:2008tp}.

\begin{figure}[H]
\centering
\begin{tabular}{cc}
\includegraphics[width=8cm, height=7cm]
{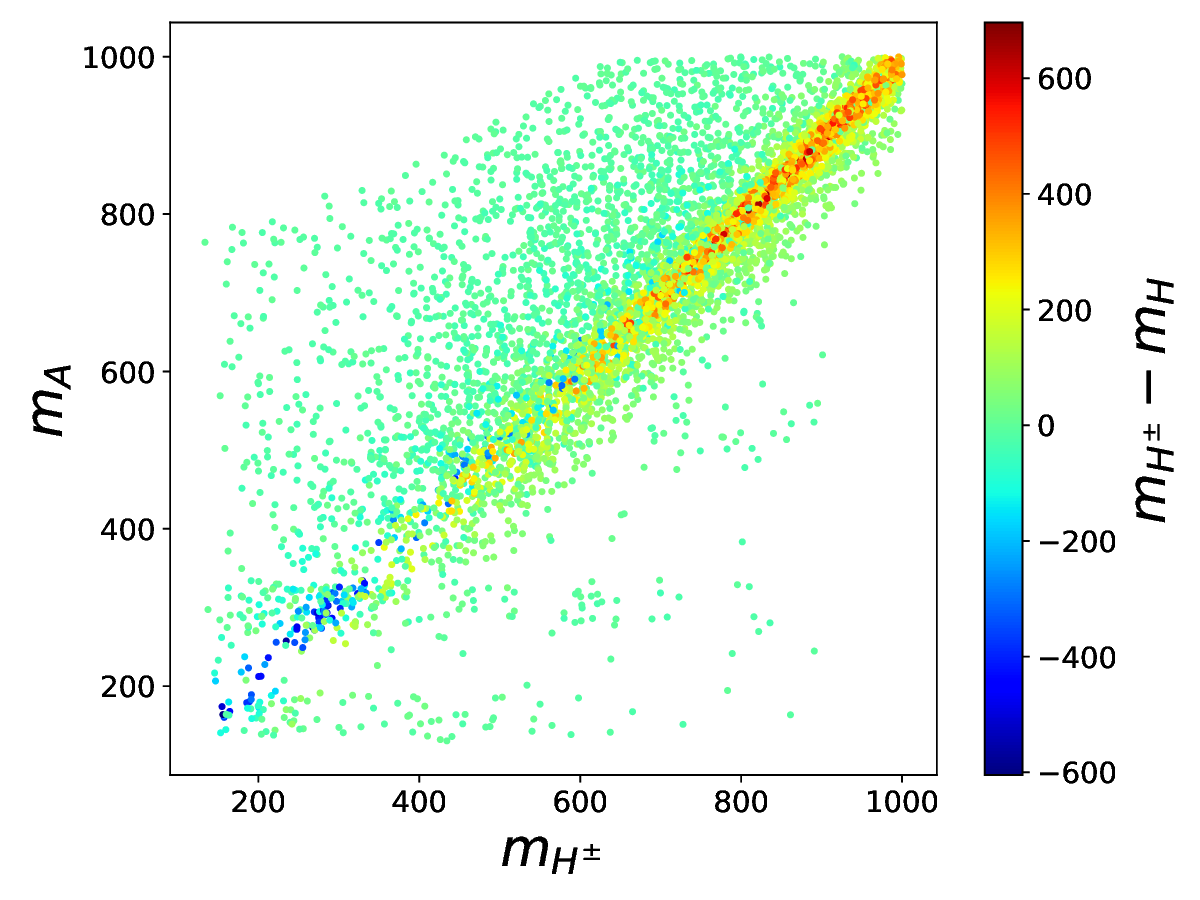}
&
\includegraphics[width=8cm, height=7cm]
{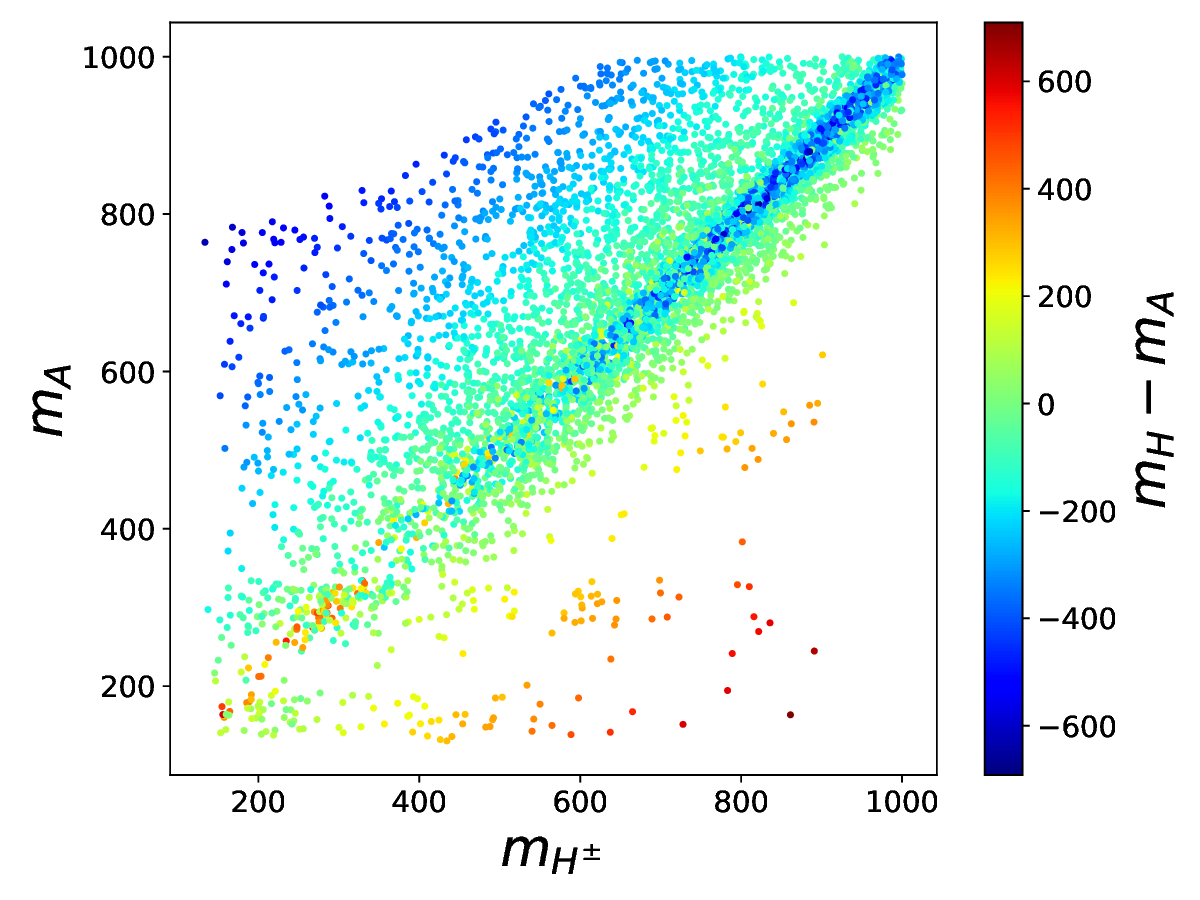}
\end{tabular}
\caption{\label{scanMAMHMHp}
The scatter plots show correlations 
in the parameter space:
$(m_A,\, m_{H^\pm},\, m_H)$ in the left panel and
$(m_A,\, m_{H^\pm},\, m_H-m_A)$ in the 
right panel.}
\end{figure}

\begin{figure}[H]
\centering
\begin{tabular}{cc}
\includegraphics[width=8cm, height=7cm]
{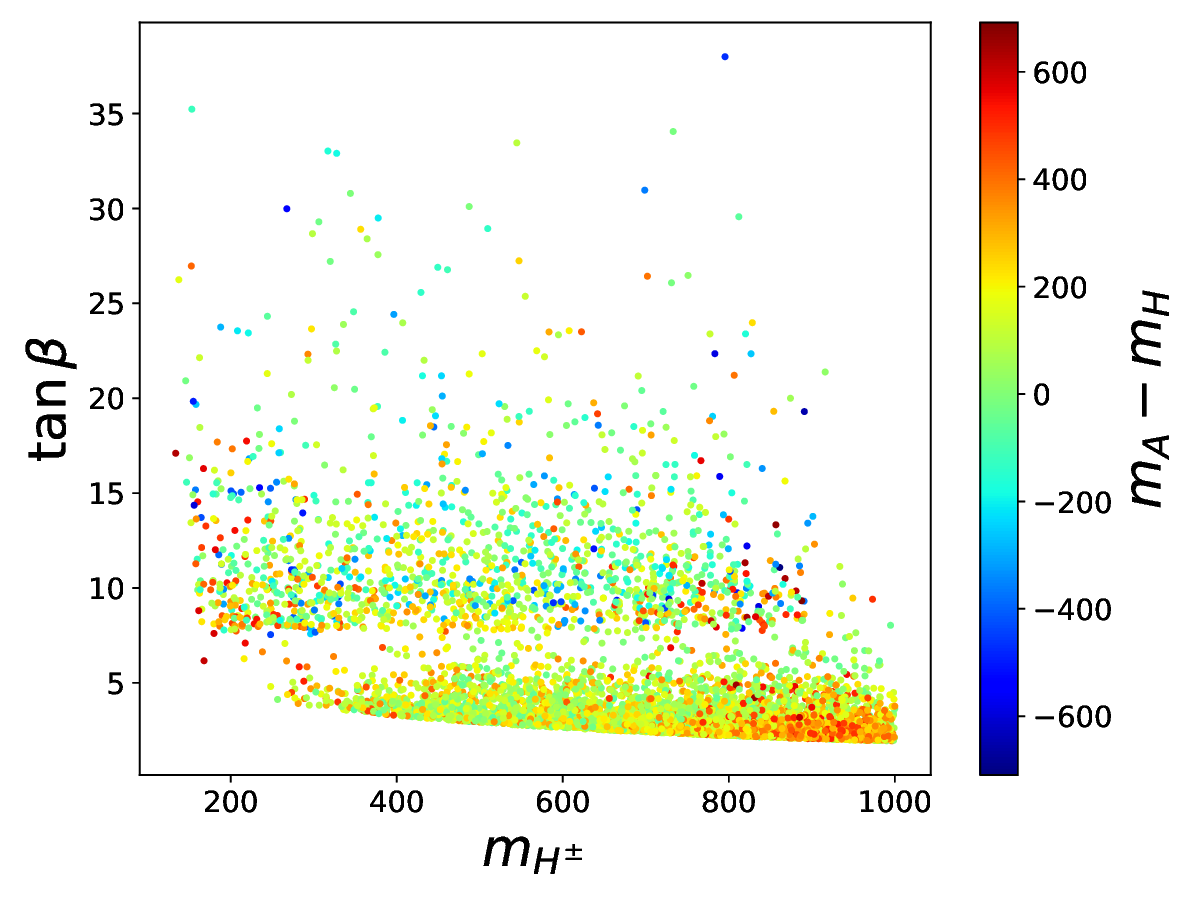}
&
\includegraphics[width=8cm, height=7cm]
{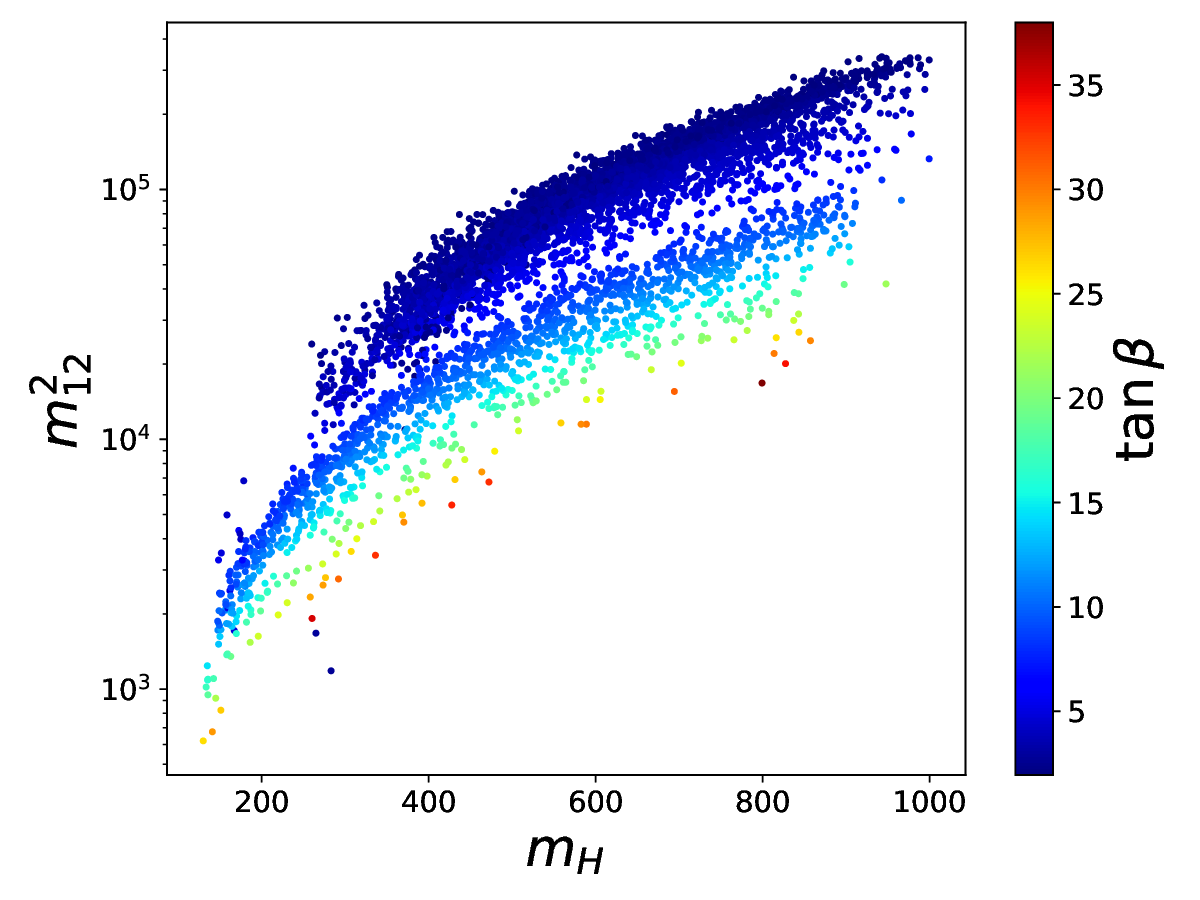}
\end{tabular}
\caption{\label{scanmMAMHpMphi1}
The scatter plots show the correlations among
the splitting mass $m_A - m_H$ with $m_{H^\pm}$
and  $\tan\beta$ for the left panel,
and among $m_{12}^2$ with $m_{H^\pm}$, and 
$\tan\beta$ for the right panel.}
\end{figure}

We now discuss the parameter space constrained by the following
scatter plots. 
In Fig.~\ref{scanMAMHMHp}, the left plot indicates 
for the correlations 
among the scalar mass splitting $m_{H^\pm}-m_H$, $m_A$, and $m_{H^\pm}$, 
while the right panel presents those among another scalar mass
splitting and $m_H-m_A$, $m_A$, and $m_{H^\pm}$.
For the left panel, when $m_A = m_{H^\pm}$, we observe that
$m_{H^\pm}-m_H$ can vary from $-600$~GeV to $600$~GeV. In the
region where $m_A > m_{H^\pm}$, the data favor the range
$-100$~GeV $\leq m_{H^\pm}-m_H \leq 100$~GeV. In the right panel,
the points with $m_A > m_H$ are more favored than the
other cases.
In Fig.~\ref{scanmMAMHpMphi1}, we show the correlations among
the splitting mass $m_A - m_H$ with $m_{H^\pm}$ and  $\tan\beta$
for the left panel,
and among $m_{12}^2$ with $m_{H^\pm}$, and $\tan\beta$ for the
right panel. The allowed regions indicate that the preferred
valid points mainly lie in the pattern $-100$~GeV $\leq m_A
- m_H \leq \sim 500$~GeV and $2 \leq \tan\beta \leq \sim 20$
in the left plot. The data shows that the
viable regions are concentrated in $10^3 \leq m_{12}^2 \leq
5 \cdot 10^5$ and $2 \leq \tan\beta \leq \sim 20$ over the
entire charged Higgs mass range in the right plot.
After obtaining the viable parameter space for 
the Type-X THDM, we transfer the allowed points to 
the program~{\tt H-COUP} version~3~\cite{Aiko:2023xui} 
to evaluate the branching ratios of the charged Higgs boson. 
All branching ratios presented in the following paragraphs
are computed using {\tt H-COUP} version~3~\cite{Aiko:2023xui}, 
which takes into account one-loop electroweak corrections 
as well as one- and two-loop QCD corrections for 
two-body decays of scalar particles; 
see~\cite{Aiko:2021can} for more details.

\begin{figure}[H]
\centering
\begin{tabular}{cc}
\includegraphics[width=8cm, height=7cm]
{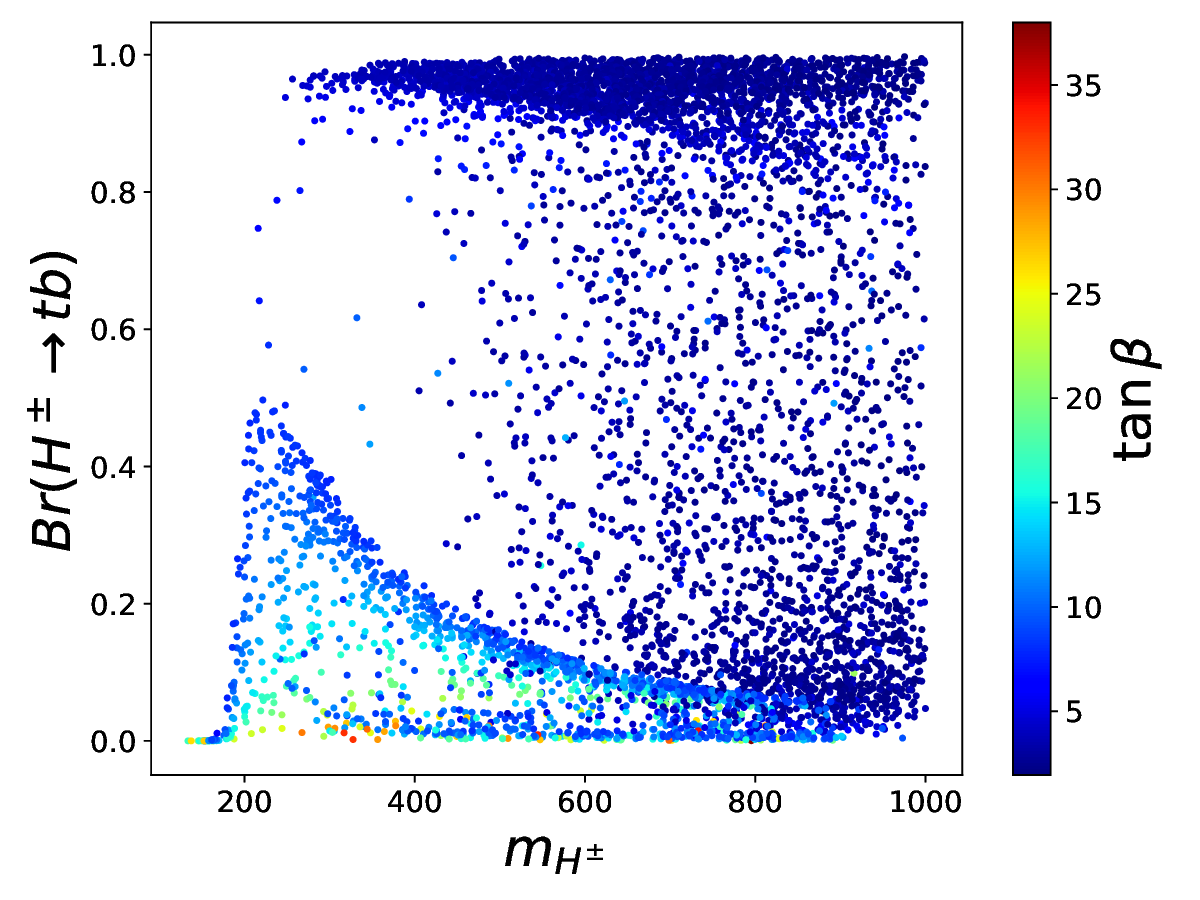}
&
\includegraphics[width=8cm, height=7cm]
{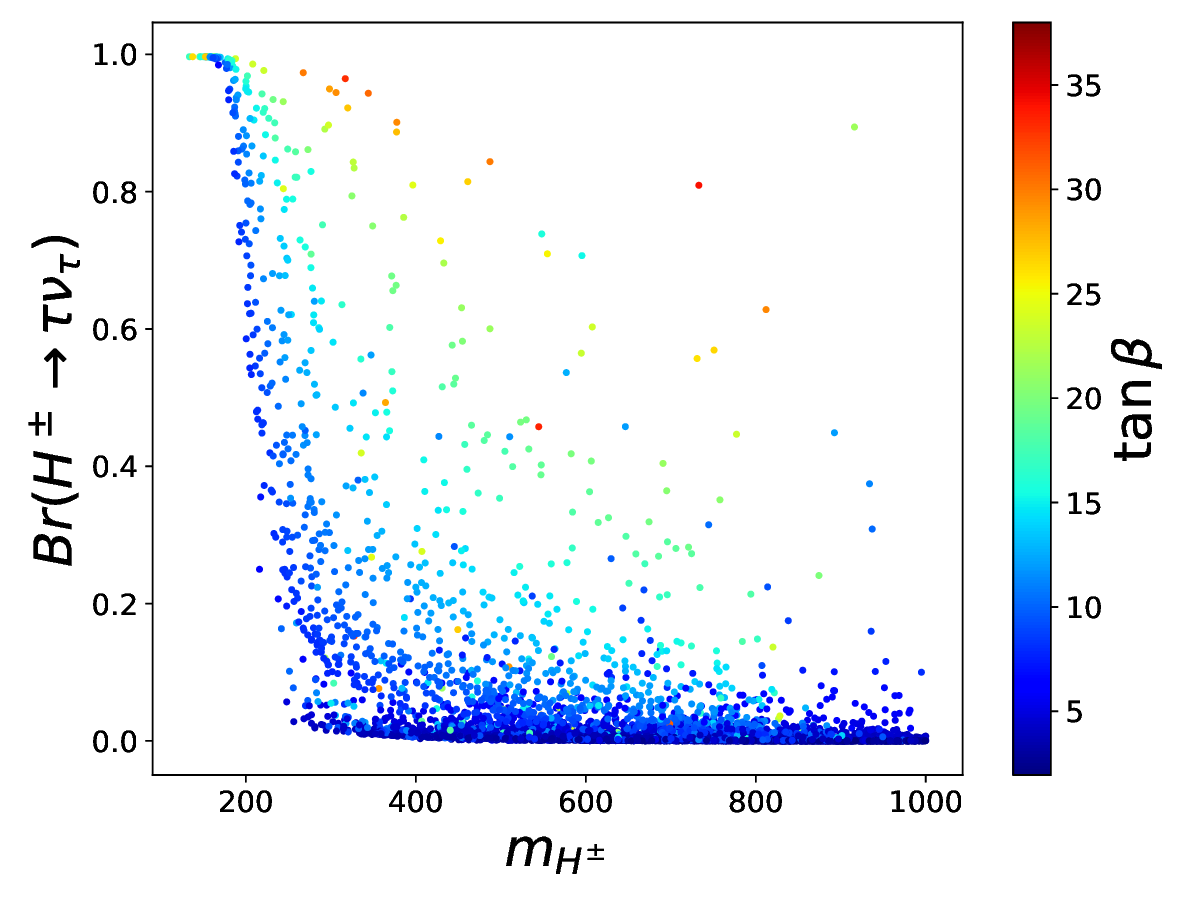}
\\
\includegraphics[width=8cm, height=7cm]
{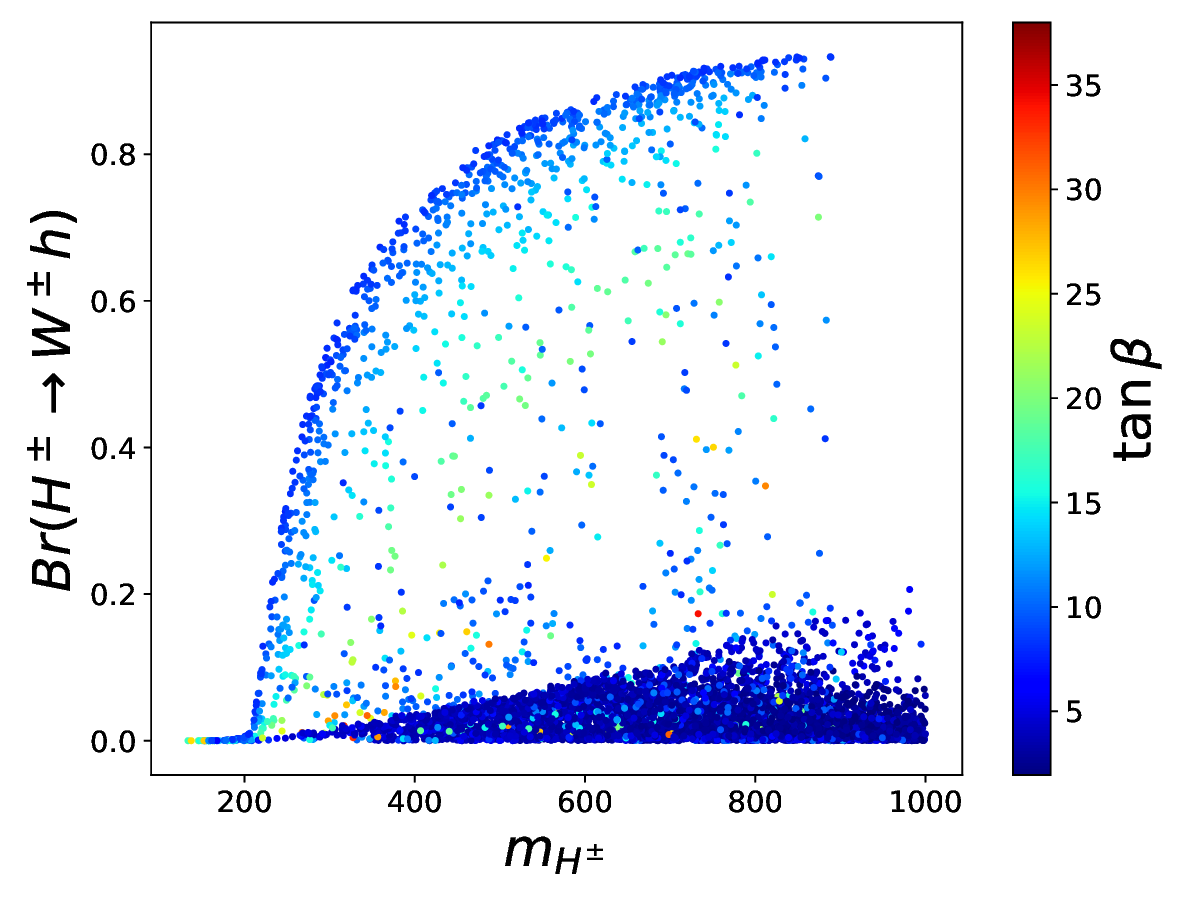}
&
\includegraphics[width=8cm, height=7cm]
{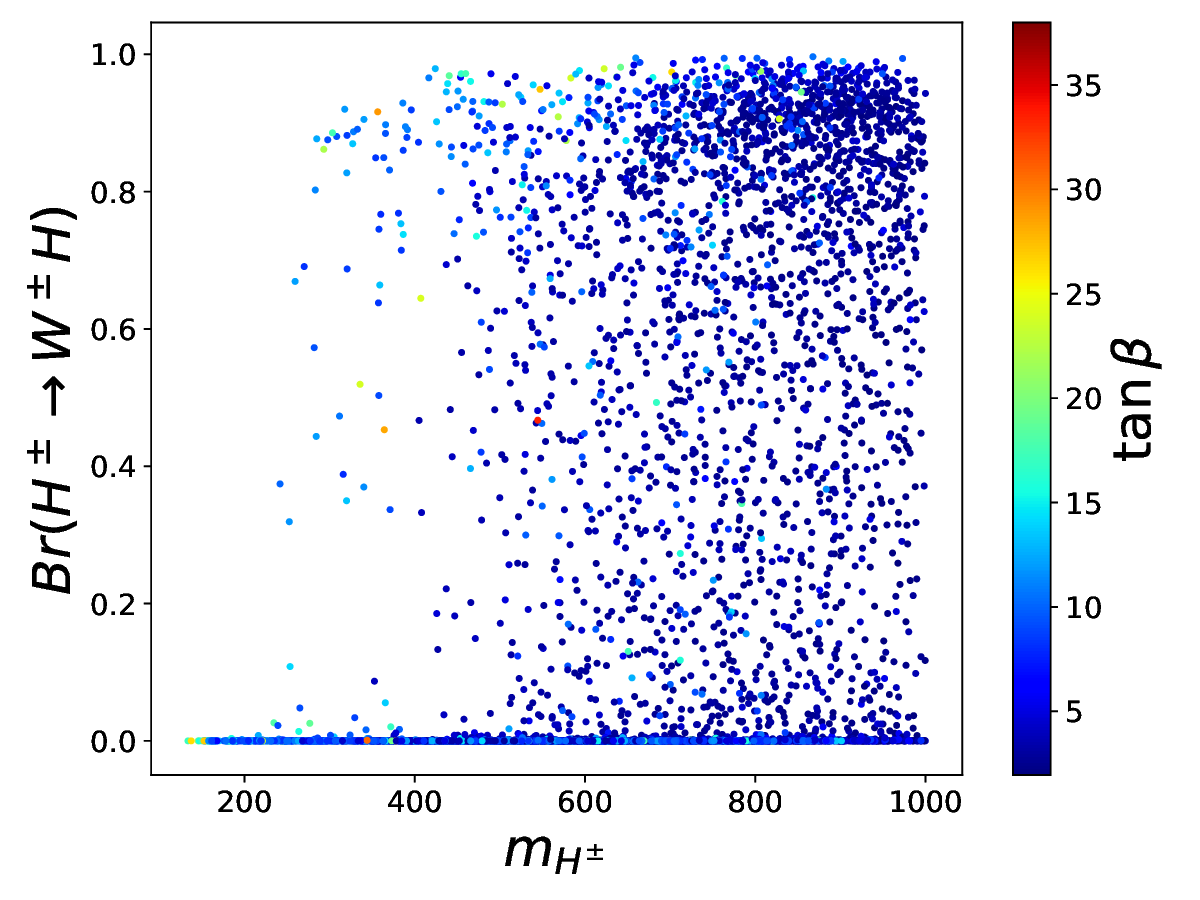}
\end{tabular}
\caption{\label{tbWphi}
Branching ratios for $H^\pm \to tb$ (left above),
$H^\pm \to \tau\nu_{\tau}$ (right above),
$H^\pm \to Wh$ (left below), and $H^\pm \to WH$
(right below) are generated within the allowed
parameter regions of the Type-X THDM. }
\end{figure}

We present the dominant decay modes of the charged Higgs
in the Type-X THDM in the following paragraphs. In the
Fig.~\ref{tbWphi}, we examine the branching
fractions of the charged Higgs boson into top and bottom
quarks (left above), $H^\pm \to \tau\nu_{\tau}$ (right above), 
$H^\pm \to Wh$ (left below) and  $H^\pm \to WH$ (right below). 
All scatter plots are
generated in the plane of $m_{H^\pm}$ and $t_{\beta}$.
Below the regions of 
$m_{H^\pm} \lesssim m_W+m_h\sim 220$ GeV, 
the decay rates $H^\pm \to \tau\nu_{\tau}$ are dominant 
when $\tan\beta \gtrsim 10$. Beyond the regions
$m_{H^\pm} \gtrsim 220$ GeV (opening the decay of 
$H^\pm \to Wh$), the branching fractions
of $H^\pm \to tb$ are dominant contributions in the
regions of $\tan\beta \lesssim 5$. While the branching
fractions of $H^\pm \to Wh$ become large contributions
in comparison with those of $H^\pm \to tb$ when
$\tan\beta \gtrsim 10$. Furthermore, we observe that
when $\tan\beta \lesssim 2$, the branching fractions
of $H^\pm \to WH$ also become dominant contributions.
This explains the reason that the branching fractions
of $H^\pm \to tb$ are small in the regions of
$\tan\beta \lesssim 2$.
Two decay modes, $H^{\pm} \to tb$ and $H^{\pm} \to Wh$,
are of interest for computing the production cross sections
in the next sections.
\section{Productions $\mu^+\mu^-         
\rightarrow H^{\pm}H^{\mp}V$             
at future multi--TeV muon colliders }          
In this section, we perform calculations for the production
processes $\mu^+\mu^- \rightarrow H^{\pm}H^{\mp}V$, where
$V \equiv \gamma,~Z$, at future muon–TeV colliders. The
computer packages {\tt FeynArts/FormCalc} are employed to
evaluate these processes. All tree-level Feynman diagrams
and their corresponding amplitudes are presented in
appendix~A. The valid parameter space obtained in the
previous section is then used to generate the cross
sections. As mentioned earlier, there are four dominant
decay channels such as $H^\pm \to tb$,
$\tau\nu_{\tau}$, $Wh$, and $WH$. In this simulation, we
focus on $H^\pm \to tb$ and $H^\pm \to Wh$ as two typical
examples.

\begin{figure}[H]
\centering
\begin{tabular}{cc}
$\mu^+\mu^- \to 
H^{\pm}H^{\mp}\gamma \to ttbb\gamma$
&
$\mu^+\mu^- \to 
H^{\pm}H^{\mp}\gamma \to ttbb\gamma$
\\
\includegraphics[width=8.5cm, height=7cm]
{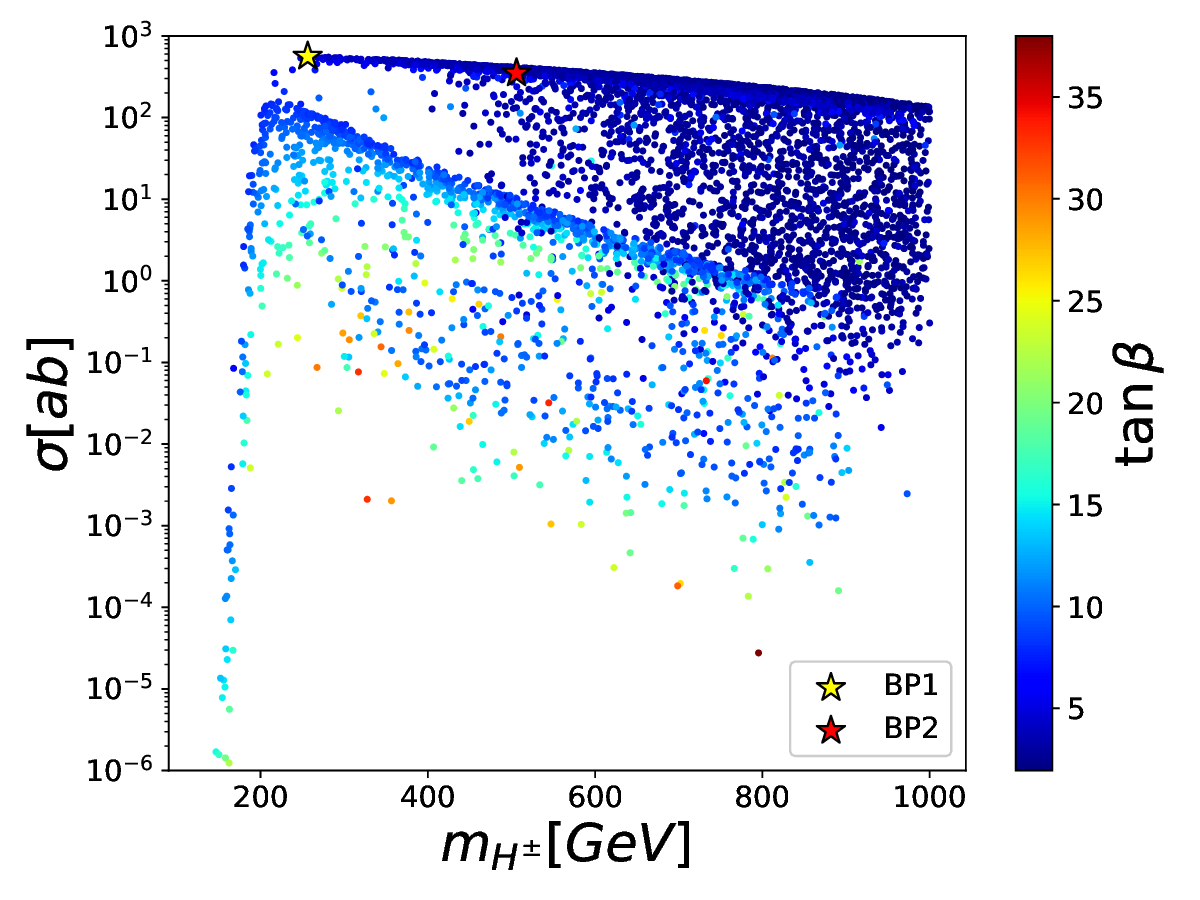}
&
\includegraphics[width=8.5cm, height=7cm]
{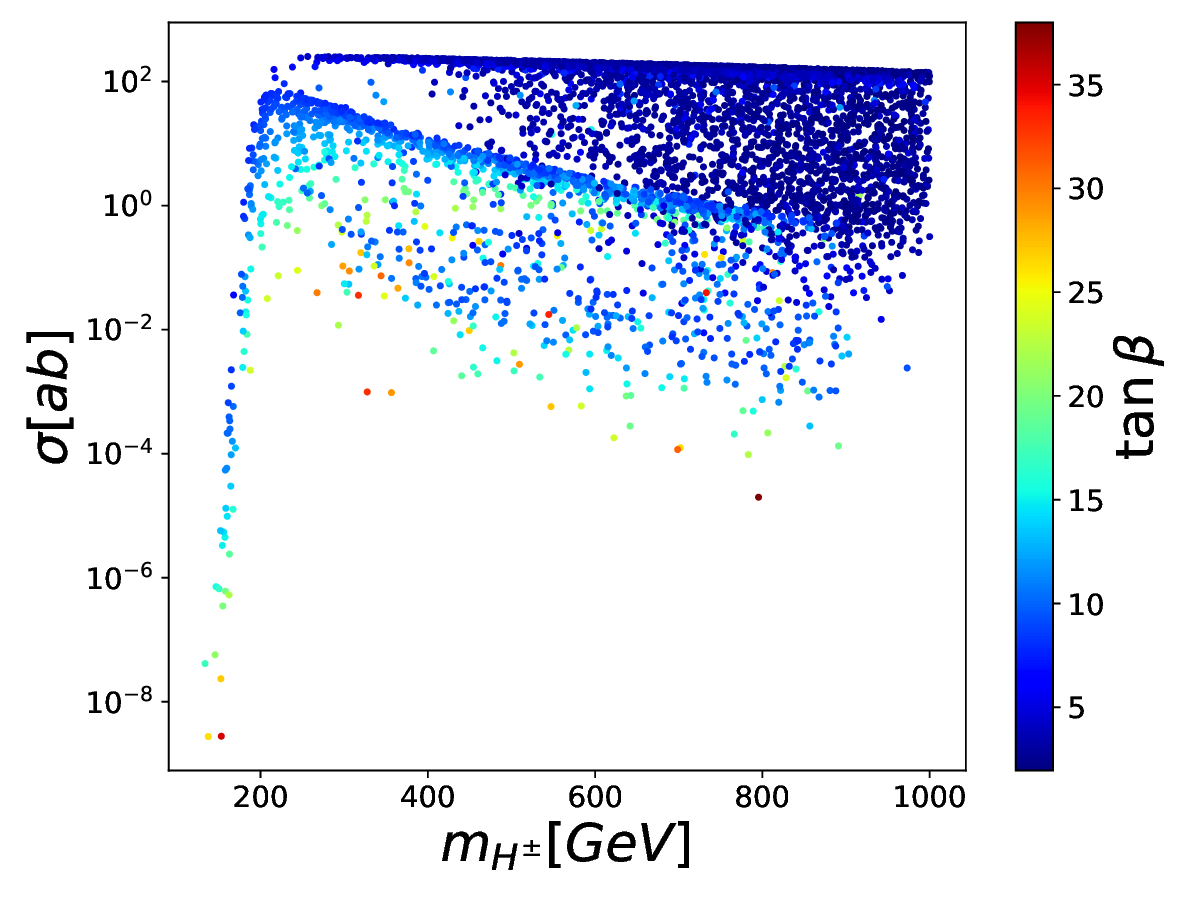}
\end{tabular}
\caption{\label{mumu2tbtbgam}
Cross-sections for the process 
$\mu^+\mu^- \to H^{\pm}H^{\mp}\gamma \to ttbb\gamma$ 
are genrated in the allowed parameter spaces of 
the Yype-X THDM at $\sqrt{s}=3$ TeV in the left panel 
and at $\sqrt{s}=5$ TeV in the right panel.
}
\end{figure}

\begin{figure}[H]
\centering
\begin{tabular}{cc}
$\mu^+\mu^- \to 
H^{\pm}H^{\mp}\gamma \to WWhh\gamma$
&
$\mu^+\mu^-\to 
H^{\pm}H^{\mp}\gamma \to WWhh\gamma$
\\
\includegraphics[width=8.5cm, height=7cm]
{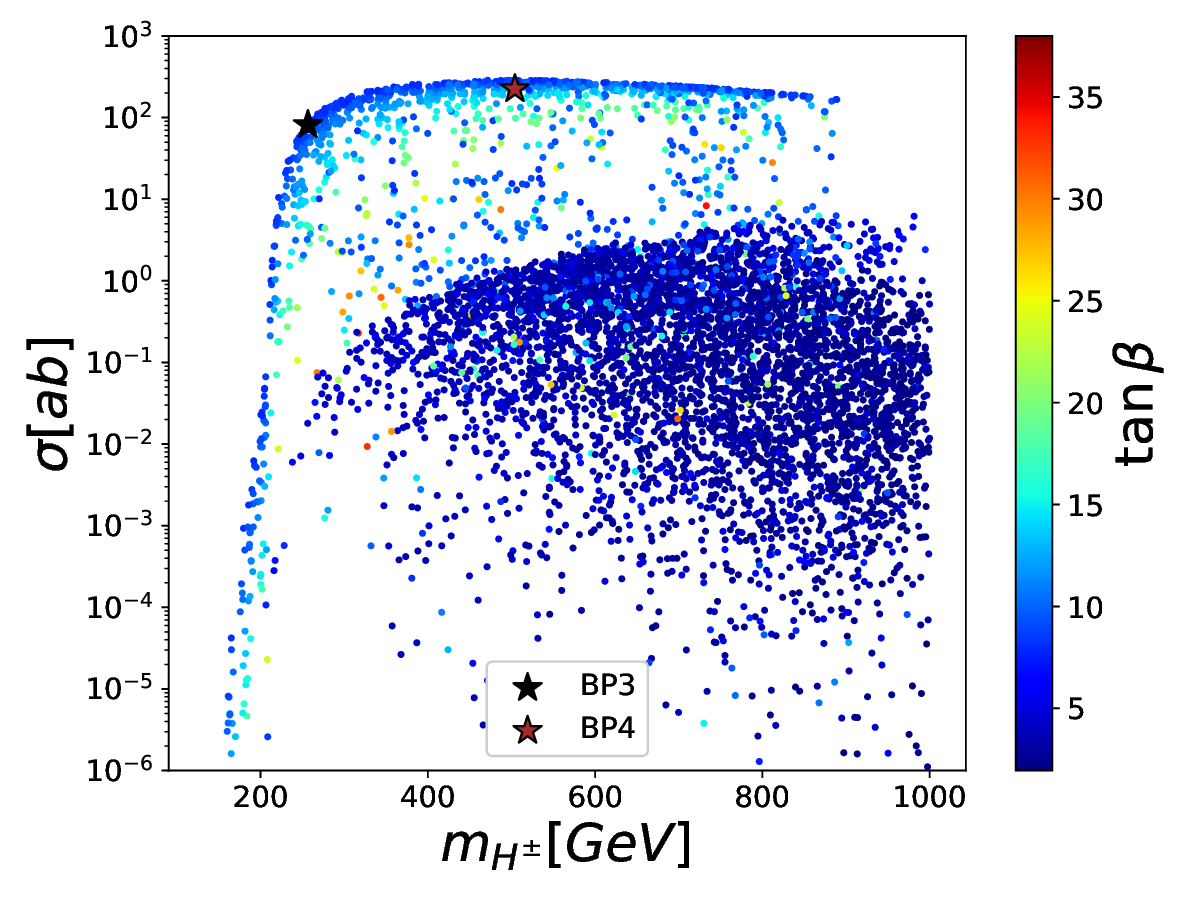}
&
\includegraphics[width=8.5cm, height=7cm]
{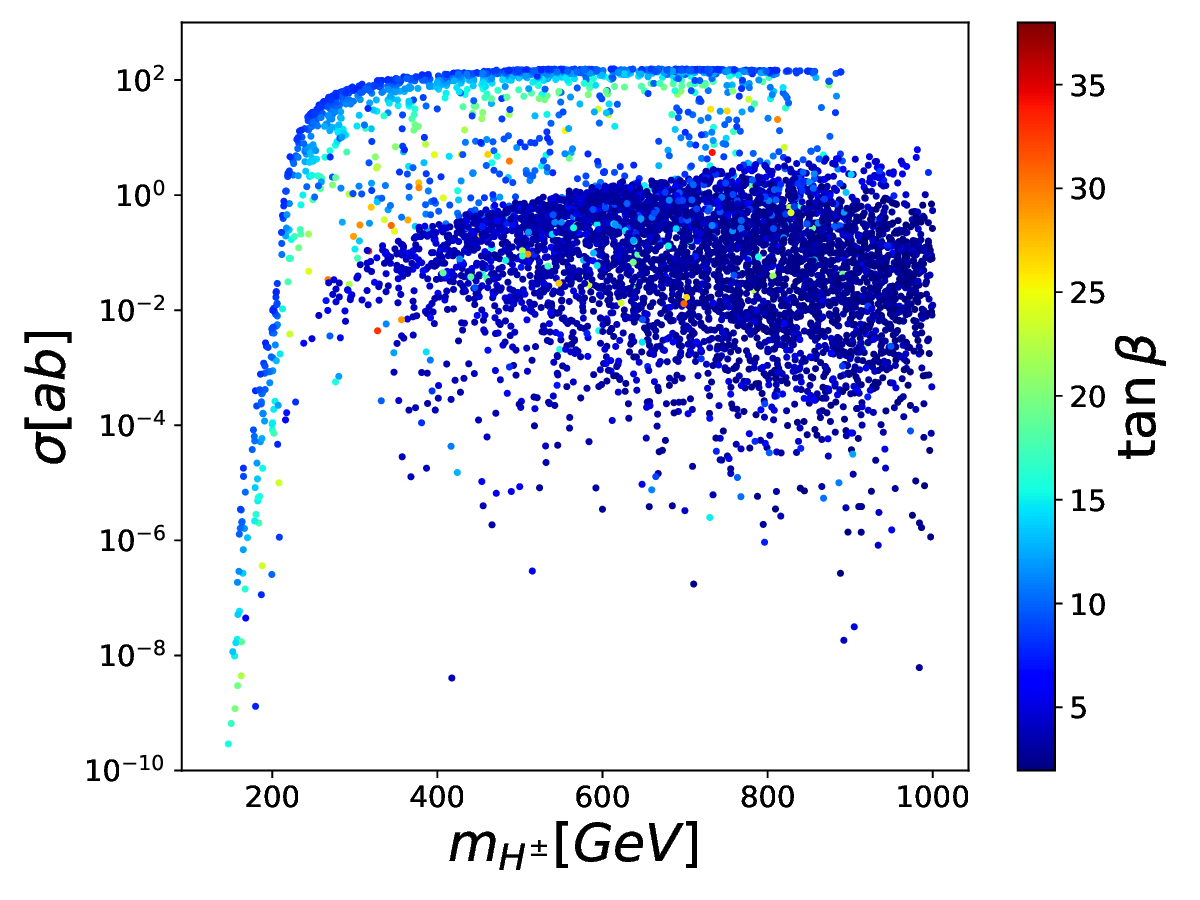}
\end{tabular}
\caption{\label{mumu2whwhgam}
Cross-sections for the process 
$\mu^+\mu^- \to H^{\pm}H^{\mp}\gamma \to WhWh\gamma$ 
are genrated in the allowed parameter spaces of
the Type-X THDM at $\sqrt{s}=3$ TeV in the left panel and 
at $\sqrt{s}=5$ TeV in the right panel.
}
\end{figure}

\begin{figure}[H]
\centering
\begin{tabular}{cc}
$\mu^+\mu^- \to 
H^{\pm}H^{\mp}Z \to ttbbZ$
&
$\mu^+\mu^- \to  
H^{\pm}H^{\mp}Z \to ttbbZ$
\\
\includegraphics[width=8.5cm, height=7cm]
{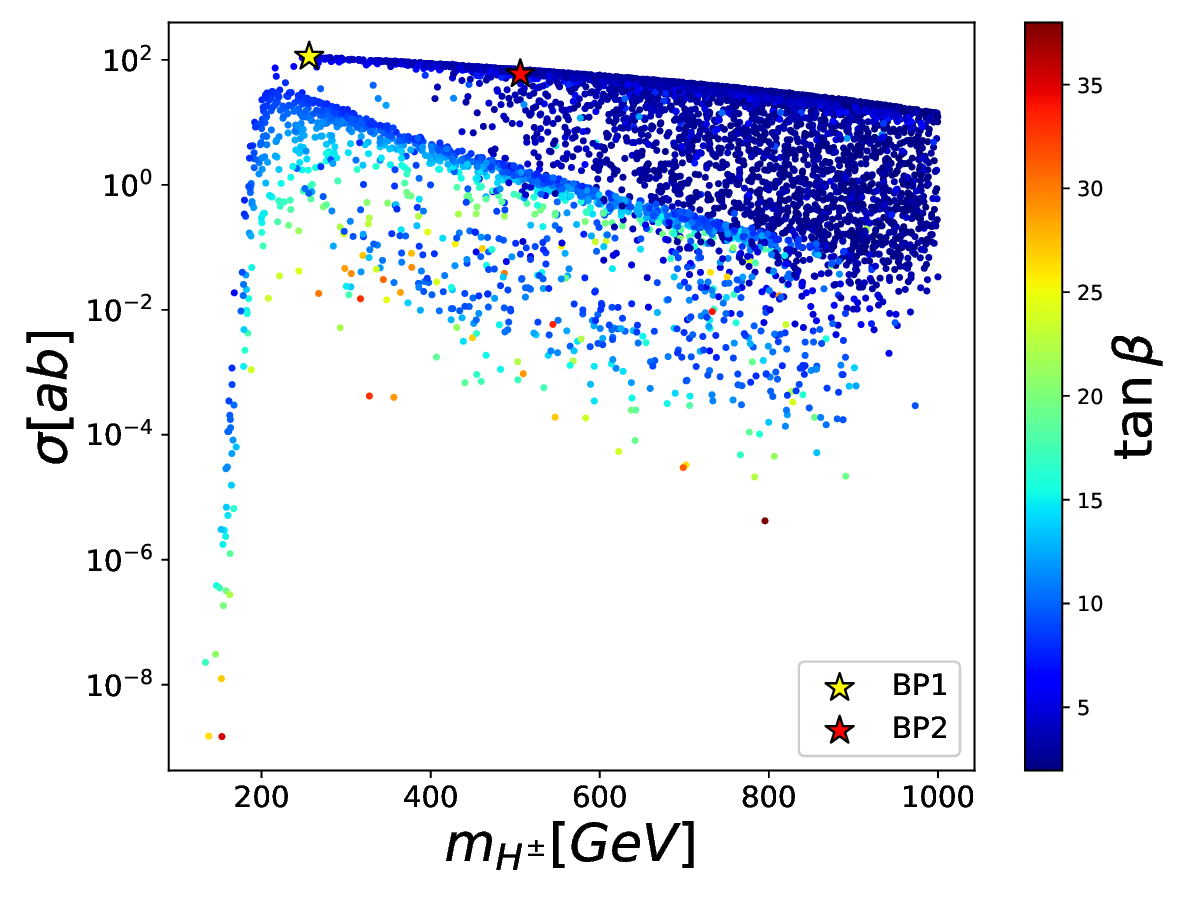}
&
\includegraphics[width=8.5cm, height=7cm]
{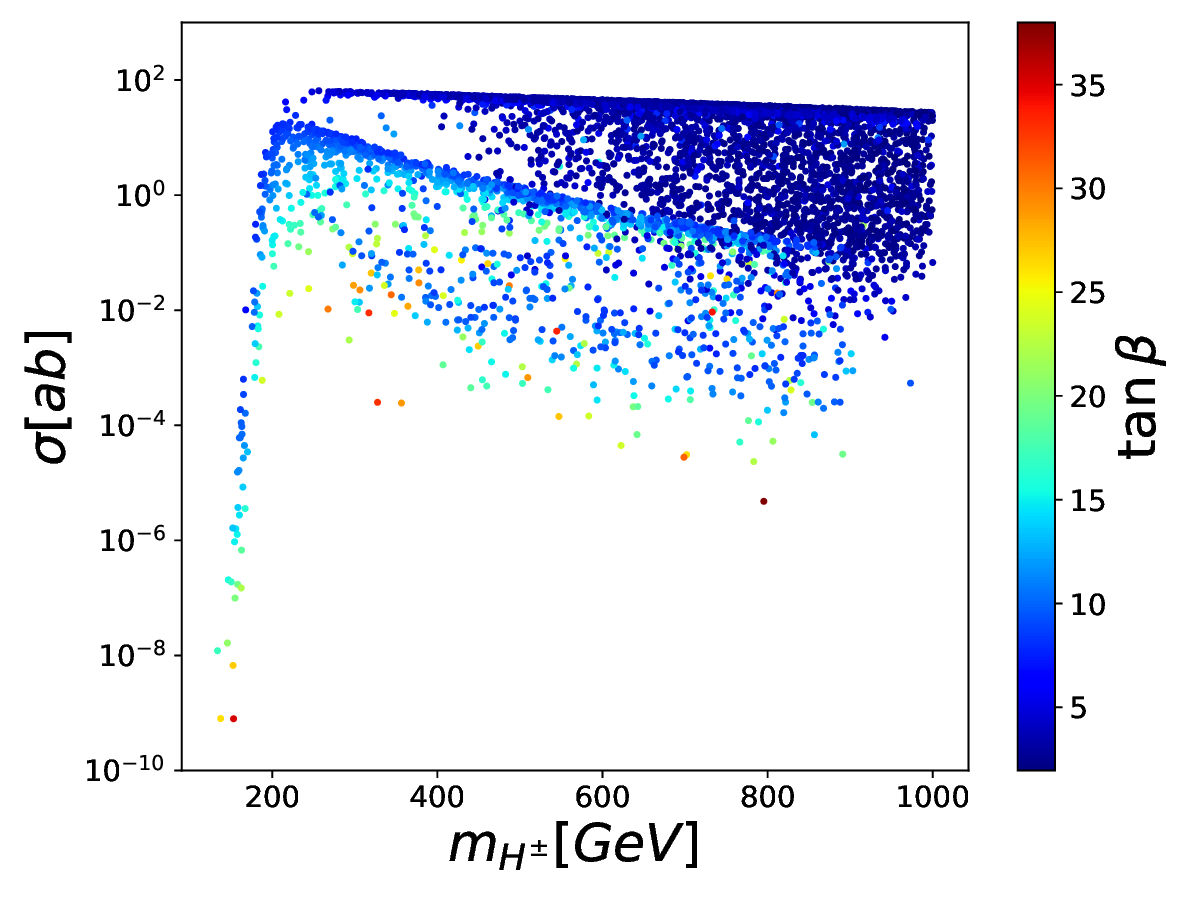}
\end{tabular}
\caption{\label{mm2tbtbZ}
Cross-sections for $\mu^+\mu^- \to 
H^{\pm}H^{\mp}Z \to ttbbZ$ are genrated in 
the allowed parameter spaces of the Type-X 
THDM at $\sqrt{s}=3$ TeV in the left panel 
and at $\sqrt{s}=5$ TeV in the right panel.
}
\end{figure}

\begin{figure}[H]
\centering
\begin{tabular}{cc}
$\mu^+\mu^- \to  
H^{\pm}H^{\mp}Z \to WWZhh$
&
$\mu^+\mu^- \to  
H^{\pm}H^{\mp}Z \to WWZhh$
\\
\includegraphics[width=8.5cm, height=7cm]
{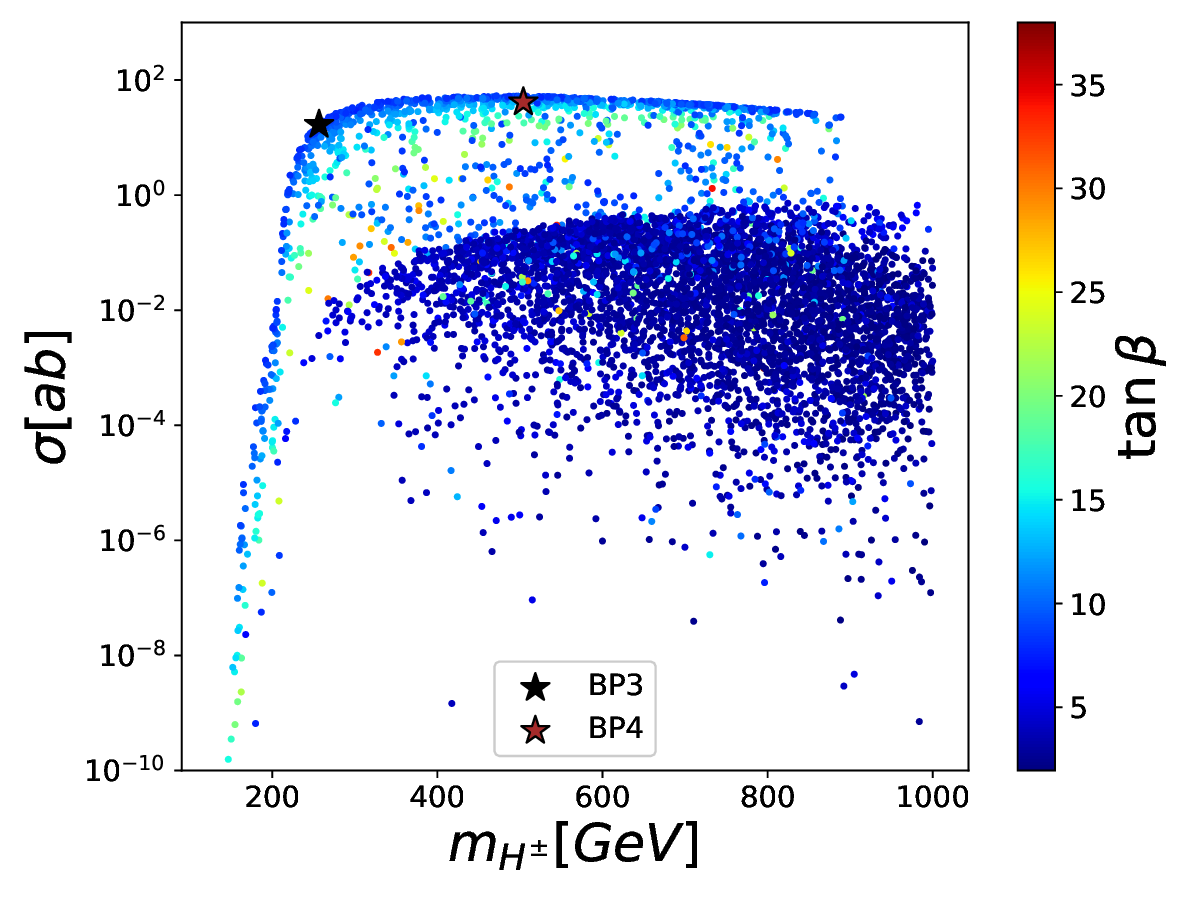}
&
\includegraphics[width=8.5cm, height=7cm]
{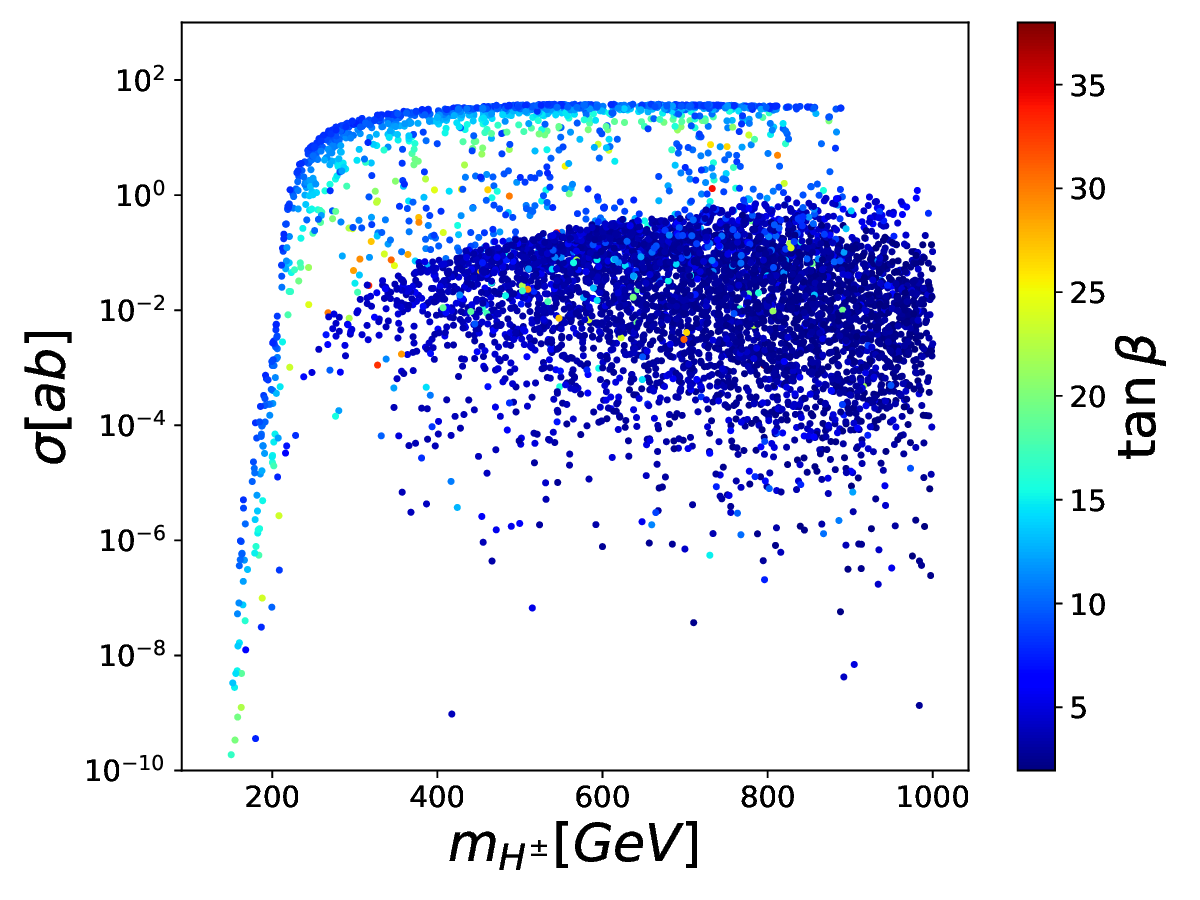}
\end{tabular}
\caption{\label{mm2whwhZ}
Cross-sections for $\mu^+\mu^- \to 
H^{\pm}H^{\mp}Z \to WWZhh$ are genrated in 
the allowed parameter spaces of Type-X THDM
at $\sqrt{s}=3$ TeV in the left panel and 
at $\sqrt{s}=5$ TeV in the right panel.}
\end{figure}

We now turn our attention to t cross section of
$\mu^+\mu^- \rightarrow H^{\pm}H^{\mp}\gamma$ at
$\sqrt{s}=3$~TeV (left panel) and $\sqrt{s}=5$~TeV (right
panel), generated within the viable parameter space obtained
above, including the subsequent decay $H^{\pm} \to tb$, as
shown in Fig.~\ref{mumu2tbtbgam}. For the results, we apply
angular and energy cuts on the external photon as follows:
$-0.98 \leq \cos\theta_{\gamma} \leq 0.98$ and $E_{\gamma}
\geq 5$~GeV. The results indicate that the cross sections peak
around $M_{H^\pm} \sim 300$~GeV. In the Type-X scenario, the
coupling of the charged Higgs to top and bottom quarks is
enhanced at small values of $\tan\beta$, which explains why
the cross sections are larger in the low-$\tan\beta$ region
compared to other regions. It is interesting to observe that
the cross section can reach the order of $\mathcal{O}(1)$~fb.
With the help of the high integrated luminosity planned 
at multi--TeV muon colliders, ranging from $500$~fb$^{-1}$ 
to $3000$~fb$^{-1}$, we can expect approximately 
$500$ to $3000$ events to be observed
at the colliders.
The cross sections for the process 
$\mu^+\mu^- \rightarrow H^{\pm}H^{\mp}\gamma \to WhWh\gamma$ 
at $\sqrt{s}=3$~TeV (left panel) and at 
$\sqrt{s}=5$~TeV (right panel), 
taking into account the sequential decay $H^{\pm} \to Wh$, 
are examined within the viable parameter space of the Type-X THDM. 
Over the entire range of charged Higgs masses, the cross sections 
exhibit similar behavior to the cases discussed above, as shown in 
Fig.~\ref{mumu2whwhgam}. We find that the cross sections peak in the 
region $M_{H^\pm} \in [250,\,400]$~GeV and are enhanced for 
$\tan\beta \gtrsim 10$. The cross sections can reach the order of 
$\mathcal{O}(0.5)$~fb. Consequently, one can expect approximately 
$250$ to $1500$ events for  $\mathcal{L}=500$~fb$^{-1}$ 
to $\mathcal{L}=3000$~fb$^{-1}$ operating 
at future multi--TeV muon colliders.
Another production channel considered in this work is 
$\mu^+\mu^- \rightarrow H^{\pm}H^{\mp}Z$. The corresponding cross 
sections at $\sqrt{s}=3$~TeV (left panel) and at $\sqrt{s}=5$~TeV 
(right panel) are generated within the viable parameter space 
obtained above, including the charged Higgs decay mode 
$H^{\pm} \to tb$, as shown in Fig.~\ref{mm2tbtbZ}. 
The cross sections for this process are about one order of magnitude 
smaller than those of $\mu^+\mu^- \rightarrow H^{\pm}H^{\mp}\gamma
\to tbtb\gamma$. 
Consequently, we can expect approximately $50$ to $300$ events 
at future multi--TeV muon colliders with the integrated luminosities
from $\mathcal{L}=500$~fb$^{-1}$ to $\mathcal{L}=3000$~fb$^{-1}$.
We then examine the production process considering 
the sequential decay $H^{\pm} \to Wh$, as shown in 
Fig.~\ref{mm2whwhZ}. Over the entire range of charged 
Higgs masses, we again find that the cross sections are 
largest in the region $250~\text{GeV} 
\leq m_{H^\pm} \leq 400~\text{GeV}$ and are enhanced 
for $\tan\beta \geq \sim 10$. 
The cross sections can be obtained the order of 
$\mathcal{O}(0.07)$~fb. As a result, one can expect 
approximately $35$ to $210$ events 
for for $\mathcal{L}=500$~fb$^{-1}$ to 
$\mathcal{L}=3000$~fb$^{-1}$
integrated at future multi--TeV muon colliders. 

\section{Productions of $H^{\pm}H^{\mp}V$ 
via $\gamma\gamma$                        
fusion at future multi--TeV muon colliders }    
At future multi--TeV muon colliders, $\gamma\gamma$-fusion is also 
promising option for porbing scalar Higgs in many 
extension of the SM. For this reason, we also compute 
the production of $H^{\pm}H^{\mp}V$ via $\gamma\gamma$- 
fusion at future multi--TeV muon colliders 
in this section. The total cross section can be 
calculated by integrating the photon splitting 
functions as follows:
\begin{equation}
\label{totalgamgam}
\sigma(s)
=
\int_{\tfrac{2m_{H^\pm}}
{\sqrt{s}}}^{x_{\text{max}}}
dz \,
\left( 2z
\int_{z^2/x_{\text{max}}}^{x_{\text{max}}}
\frac{dx}{x} \, f_{\gamma/\mu}(x)
\, f_{\gamma/\mu}
\left(z^2/x\right)
\right)
\, \hat{\sigma}(\hat{s} = z^2 s).
\end{equation}
Here, the photon structure function
$f_{\gamma/\mu}(x)$ is used, with $x$
denoting the energy fraction of the
photon emitted by the incoming lepton.
The explicit formulas for $f_{\gamma/\mu}(x)$
are given in~\cite{Zarnecki:2002qr}.
In the master formulas, we adopt
$x_{\rm max} = 0.83$ as in~\cite{Telnov:1989sd}.

\begin{figure}[H]
\centering
\begin{tabular}{cc}
$\mu^+\mu^- \to \gamma\gamma \to 
H^{\pm}H^{\mp}\gamma \to ttbb\gamma$
&
$\mu^+\mu^- \to \gamma\gamma \to 
H^{\pm}H^{\mp}\gamma \to ttbb\gamma$
\\
\includegraphics[width=8.5cm, height=7cm]
{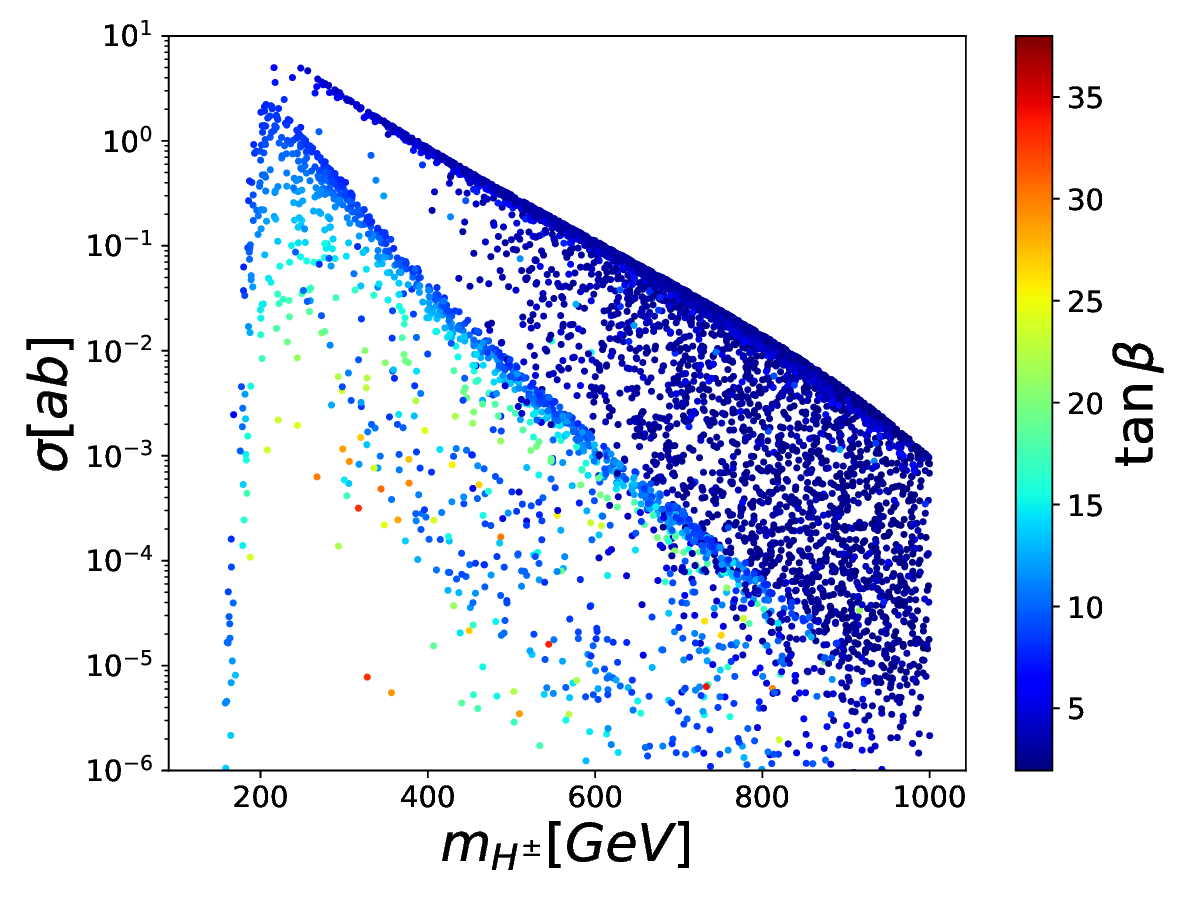}
&
\includegraphics[width=8.5cm, height=7cm]
{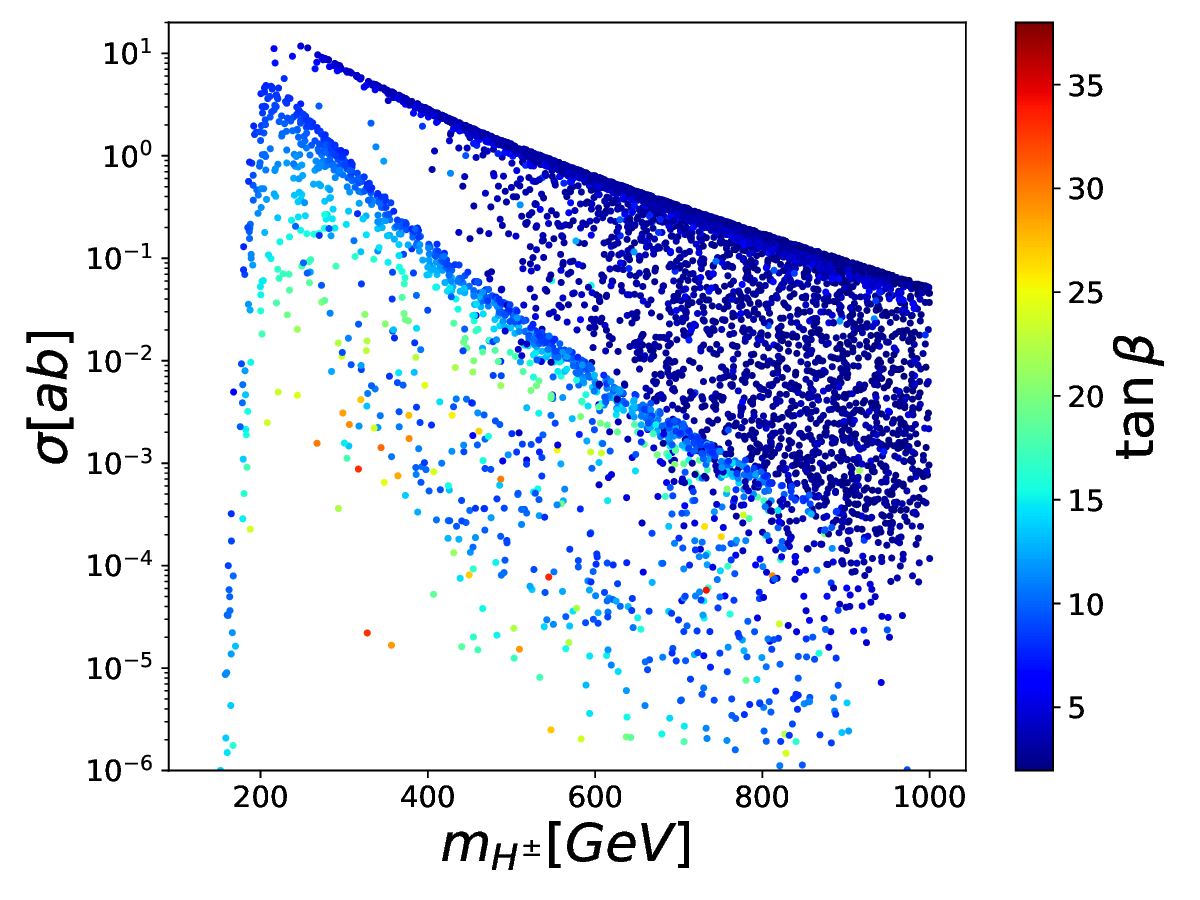}
\end{tabular}
\caption{\label{gg2tbtbgam}
Cross-sections for $\mu^+\mu^-\to \gamma\gamma
\to H^{\pm}H^{\mp}\gamma \to tbtb\gamma$ 
are genrated in 
the allowed parameter spaces of Type-X 
THDM at $\sqrt{s}=3$ TeV in the left panel 
and at $\sqrt{s}=5$ TeV in the right panel.}
\end{figure}

\begin{figure}[H]
\centering
\begin{tabular}{cc}
$\mu^+\mu^- \to \gamma\gamma \to 
H^{\pm}H^{\mp}\gamma \to WWhh\gamma$
&
$\mu^+\mu^- \to \gamma\gamma \to 
H^{\pm}H^{\mp}\gamma \to WWhh\gamma$
\\
\includegraphics[width=8.5cm, height=7cm]
{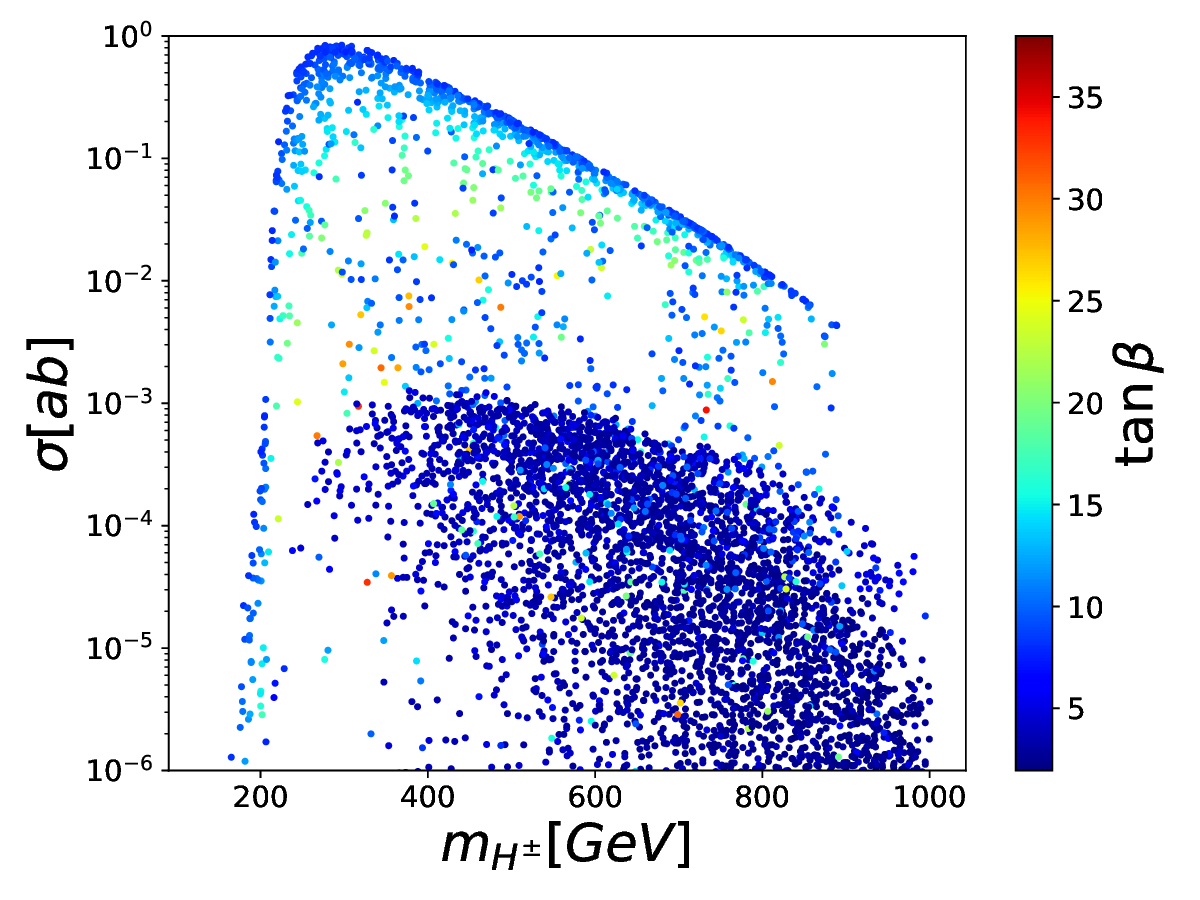}
&
\includegraphics[width=8.5cm, height=7cm]
{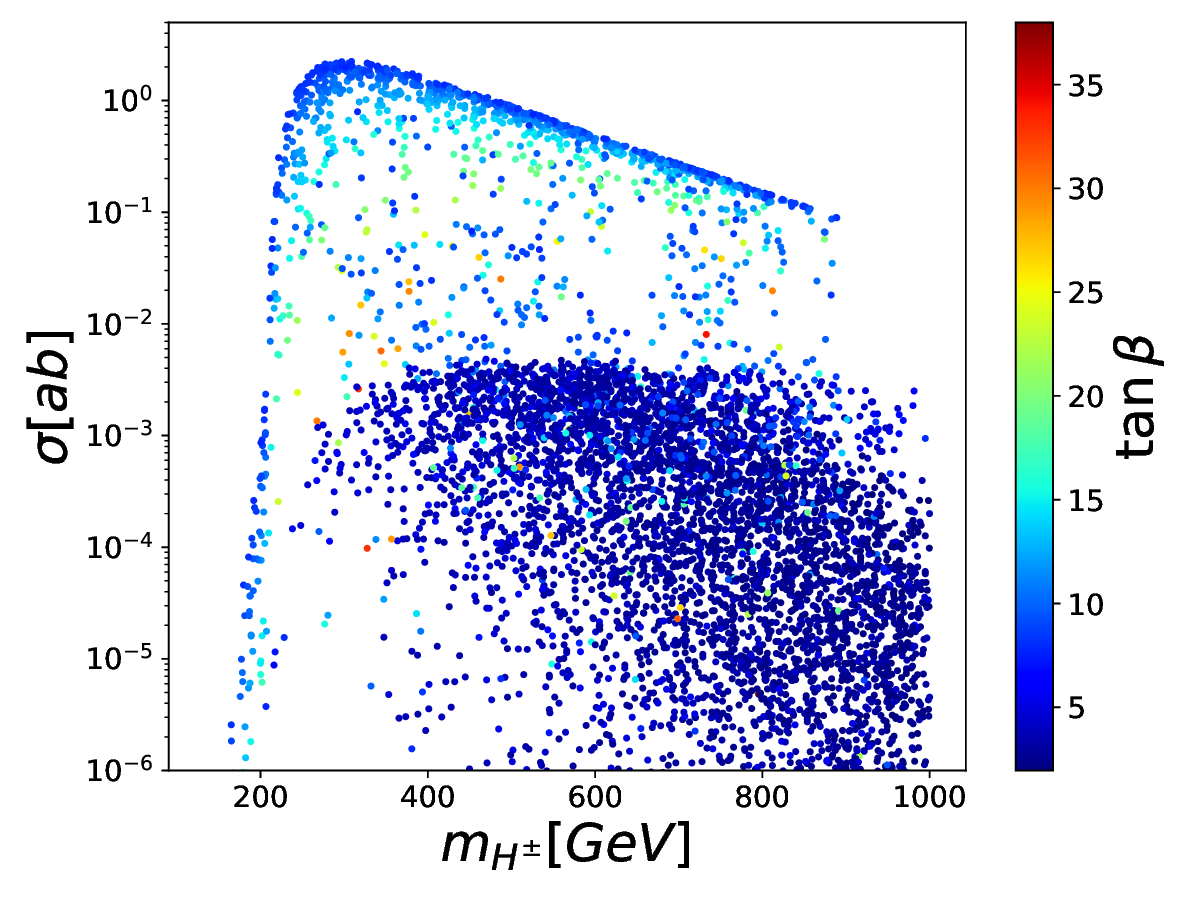}
\end{tabular}
\caption{\label{gg2whwhgam}
Cross-sections for $\mu^+\mu^-\to \gamma\gamma
\to H^{\pm}H^{\mp}\gamma \to WWhh \gamma$ 
are genrated in 
the allowed parameter spaces of THDM type-X 
for $\sqrt{s}=3$ TeV 
in the left panel and for $\sqrt{s}=5$ TeV
in the right panel.}
\end{figure}

\begin{figure}[H]
\centering
\begin{tabular}{cc}
$\mu^+\mu^- \to \gamma\gamma \to 
H^{\pm}H^{\mp}Z \to ttbbZ$
&
$\mu^+\mu^- \to \gamma\gamma \to 
H^{\pm}H^{\mp}Z \to ttbbZ$
\\
\includegraphics[width=8.5cm, height=7cm]
{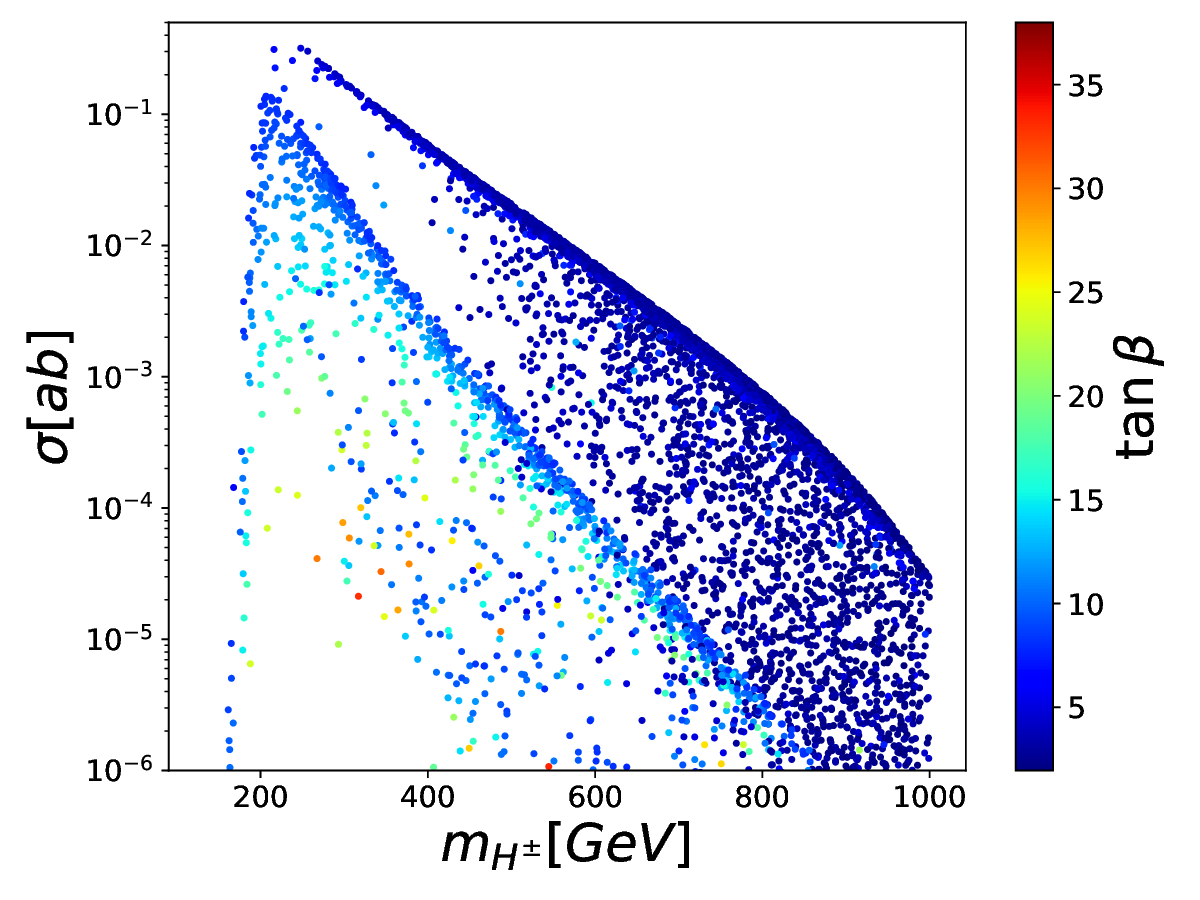}
&
\includegraphics[width=8.5cm, height=7cm]
{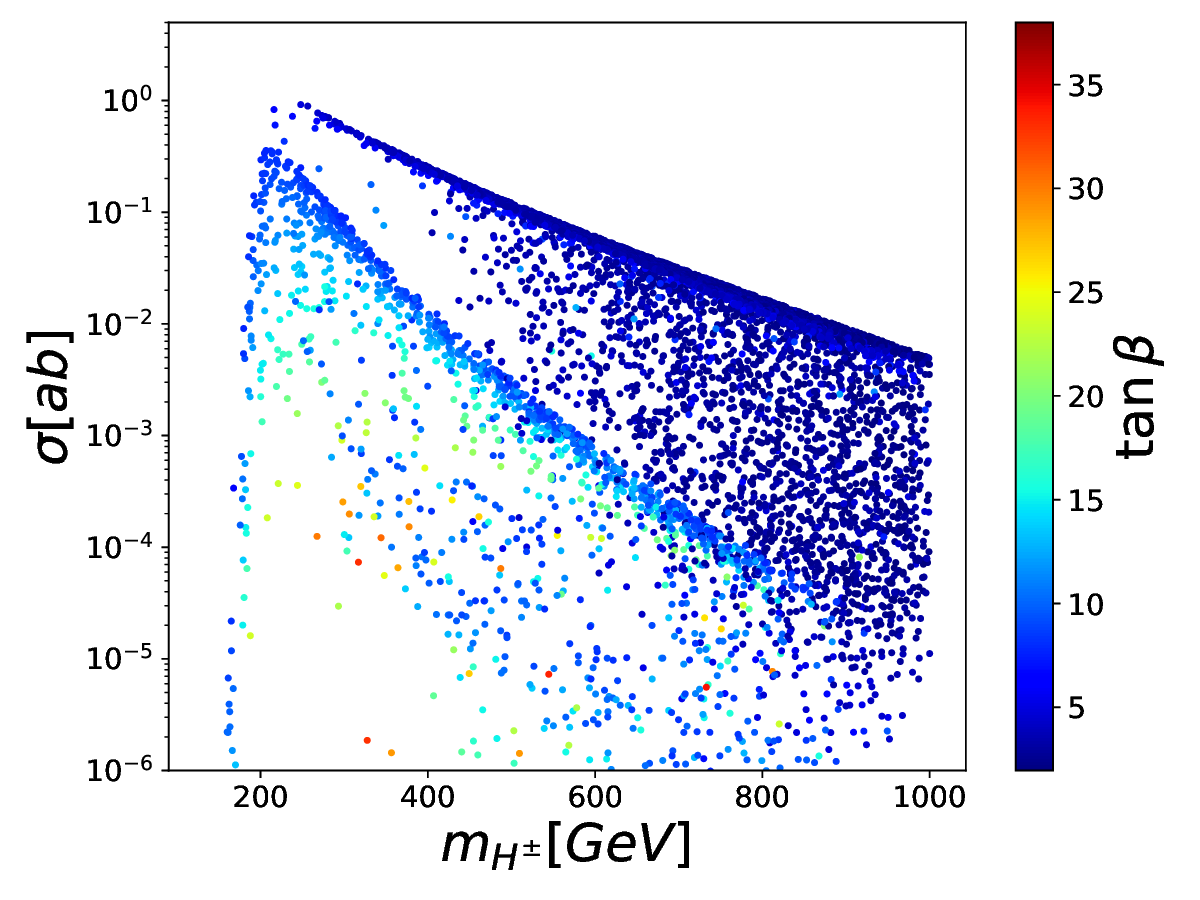}
\end{tabular}
\caption{\label{gg2tbtbZ}
Cross-sections for $\mu^+\mu^- \to \gamma\gamma \to 
H^{\pm}H^{\mp}Z \to tbtb Z$ are genrated in 
the allowed parameter spaces of THDM type-X 
for $\sqrt{s}=3$ TeV 
in the left panel and for $\sqrt{s}=5$ TeV
in the right panel.
}
\end{figure}

\begin{figure}[H]
\centering
\begin{tabular}{cc}
$\mu^+\mu^- \to \gamma\gamma \to 
H^{\pm}H^{\mp}Z \to WWhhZ$
&
$\mu^+\mu^- \to \gamma\gamma \to 
H^{\pm}H^{\mp}Z \to WWhhZ$
\\
\includegraphics[width=8.5cm, height=7cm]
{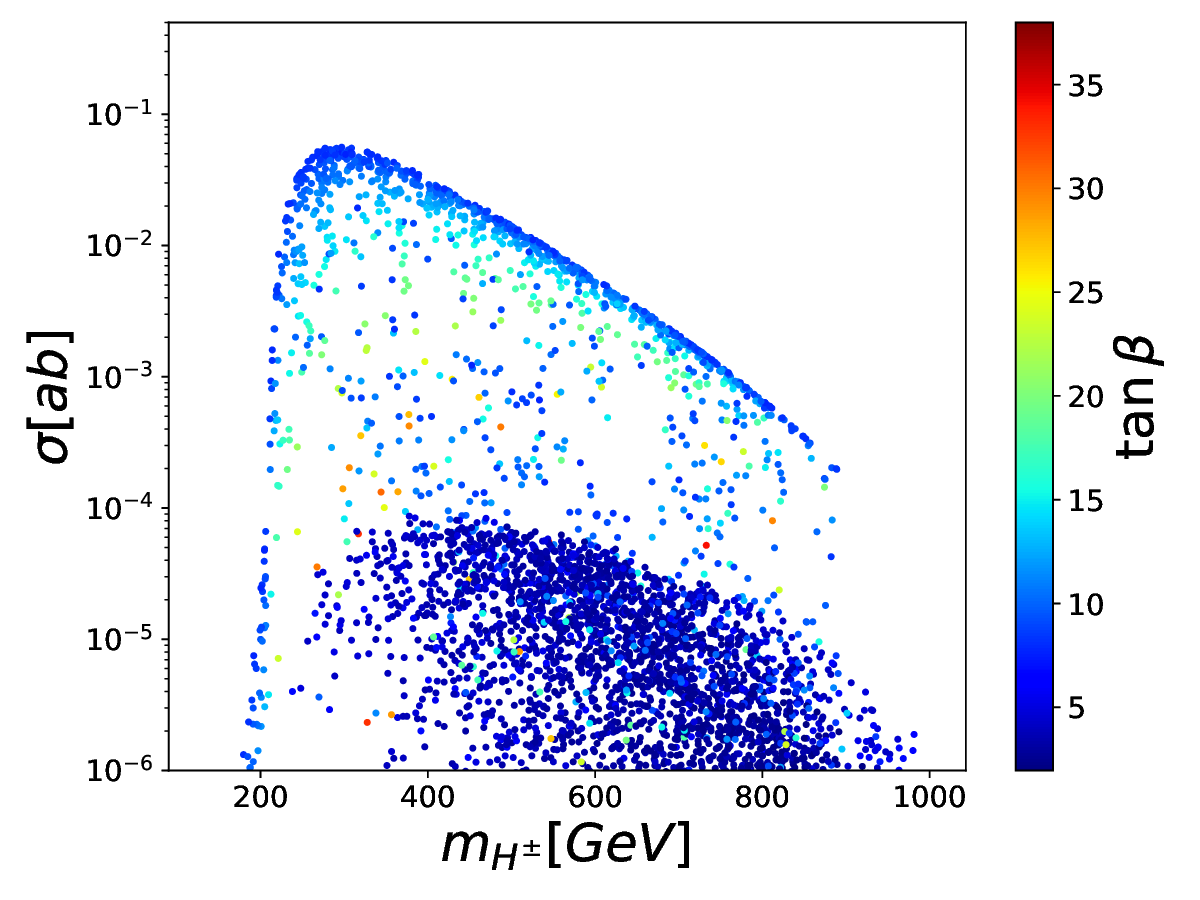}
&
\includegraphics[width=8.5cm, height=7cm]
{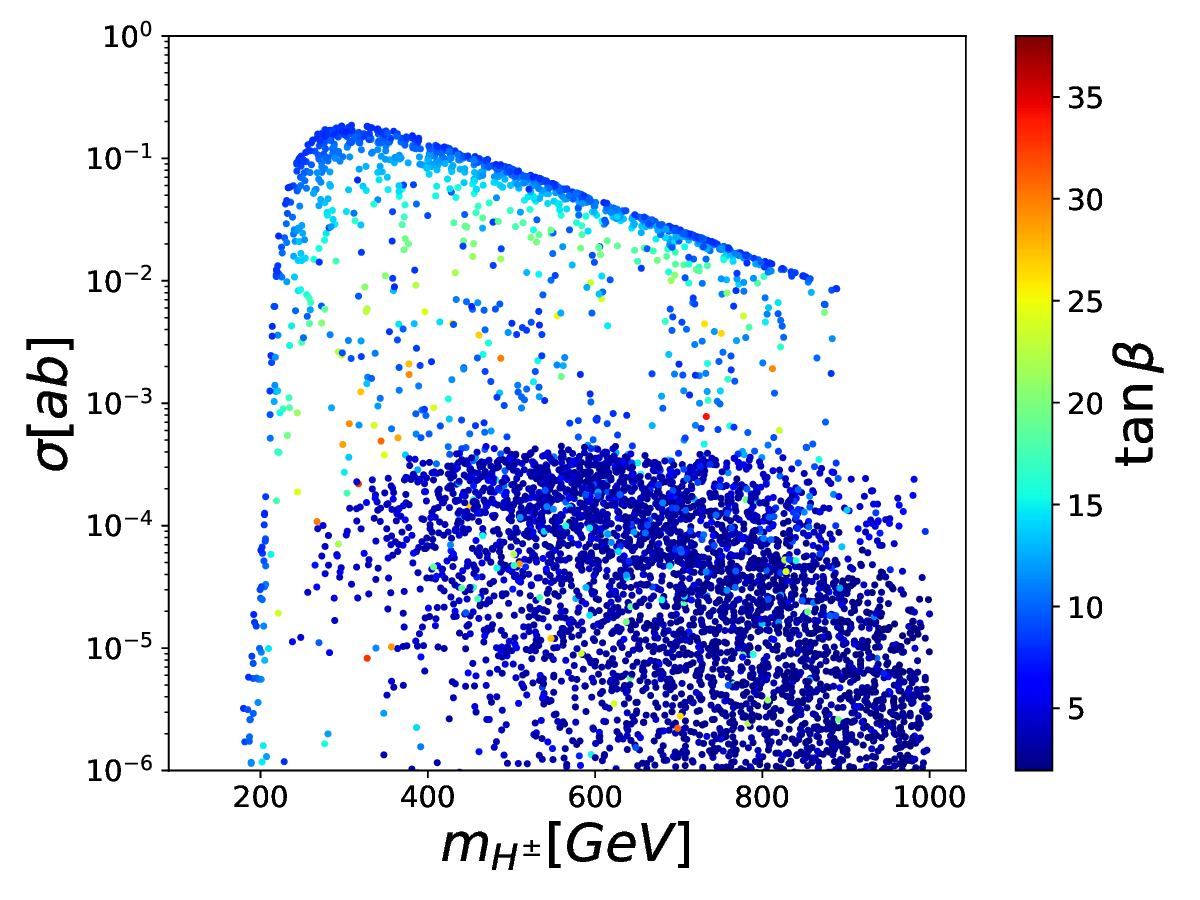}
\end{tabular}
\caption{\label{gg2whwhZ}
Cross-sections for $\mu^+\mu^- \to \gamma\gamma \to 
H^{\pm}H^{\mp}Z \to WWZhh$ are genrated in 
the allowed parameter spaces of THDM type-X 
for $\sqrt{s}=3$ TeV 
in the left panel and for $\sqrt{s}=5$ TeV
in the right panel.}
\end{figure}

We first study the cross section for 
$\mu^+\mu^- \to \gamma\gamma \to H^{\pm}H^{\mp}\gamma$ 
at $\sqrt{s} = 3$~TeV (left panel) and at $\sqrt{s} = 5$~TeV 
(right panel) within the allowed parameter space obtained 
above for the Type-X THDM. In Fig.~\ref{gg2tbtbgam}, 
the cross sections are included the sequential decay 
$H^\pm \to tb$. We have used the same cuts on external 
photon as $-0.98 \leq \cos\theta_{\gamma} \leq 0.98$ 
and $E_{\gamma} \geq 5$~GeV. The cross sections peak around 
$m_{H^\pm} \sim 300$~GeV. Since there are no diagrams with 
direct couplings of the charged Higgs boson to muons or their 
neutrinos, the cross sections in this process are inversely 
proportional to $\tan\beta$. This explains the different 
behavior of the cross sections in this process compared to 
those of $\mu^+\mu^- \to H^{\pm}H^{\mp}\gamma \to tbtb\gamma$.
Around the peak regions, the cross sections can reach the order of 
$\mathcal{O}(1)$~ab. With considering the projected integrated 
luminosity of $10$~ab$^{-1}$, only a few events are expected 
to be produced via $\gamma\gamma$-fusion at 
future multi--TeV muon colliders.
The production process 
$\mu^+\mu^- \to \gamma\gamma \to H^{\pm}H^{\mp}
\gamma \to WWhh\gamma$ 
is investigated within the parameter space 
of the Type-X THDM , as shown in Fig.~\ref{gg2whwhgam}. 
In this channel, the cross sections are rather small. 
As a result, only a very small number of events are 
expected to be detected via $\gamma\gamma$-fusion at 
future multi--TeV muon colliders.
Other production cross sections
for $\mu^+\mu^- \to \gamma\gamma \to H^{\pm}H^{\mp}Z$
at $\sqrt{s} = 3$~TeV (on the left panel)
and at (on the right panel)
are also computed in the parameter space 
of the Type-X THDM. The results are presented together 
with the branching ratios of decay modes
such as $H^\pm \to tb,~Wh$. In particular, the results
are shown for the decay mode 
$H^{\pm} \to tb$ in Fig.~\ref{gg2tbtbZ} and $H^{\pm} \to Wh$
in Fig.~\ref{gg2whwhZ}. The cross section 
are also very small. We expect to get few events of this 
proceses at future multi--TeV muon colliders. 
\section{Signal significance for $H^\pm H^\mp V$ 
at future multi--TeV muon colliders }                 
In this section, we compute the signal significances for 
$H^\pm H^\mp V$ production at future multi--TeV muon colliders.
Since the cross sections for $\mu^+\mu^- \to H^\pm H^\mp V$ 
are greater than the corresponding ones 
through $\gamma\gamma$-fusion. Additionally, both decay modes
$H^\pm \to tb$ and $H^\pm \to Wh$ are dominant in comparison 
with the other decay channels of the charged Higgs. 
For these reasons, 
we are interested in evaluating the significances
of $\mu^+\mu^- \to H^\pm H^\mp V \to tbtbV,~WhWhV$ in this 
section. The significances are evaluated for several benchmark points, 
marked as star symbols in the figures presented in the previous 
sections. These benchmark points are chosen from the regions 
where the cross sections are relatively large. 
The four selected benchmark points are summarized in Table below~\ref{BMpoints}: 
\begin{table}[ht!]
\centering
\renewcommand{\arraystretch}{1.3}
\begin{tabular}{|c|ccccccc|}
\hline\hline
BP & $M_H$ & $\Gamma_H^{\textrm{Total}}$ 
& $M_A$ & $M_{H^\pm}$ & $s_{\beta-\alpha}$ 
& $\tan\beta$ & $m_{12}^2$ \\
\hline\hline
BP1 & 342.348 & 0.1538 & 348.134 & 256.356 & 0.9992   & 4.12749  & 23726.8 \\
\hline
BP2 & 551.633 & 0.3839 & 678.245 & 506.019 & 0.999909 & 6.45338  & 44594.2 \\
\hline
BP3 & 261.685 & 0.2755 & 731.413 & 256.640 & 0.970965 & 8.01807  & 7925.15 \\
\hline
BP4 & 531.033 & 6.6749 & 619.362 & 504.104 & 0.988142 & 12.9283  & 21501.0 \\
\hline\hline
\end{tabular}
\caption{
\label{BMpoints}
Benchmark points used in the analysis. 
Masses and decay widths are in GeV.}
\end{table}

Before presenting the signal significances for 
$H^\pm H^\mp V$ production at future multi--TeV muon colliders, 
we know that the initial-state radiation (ISR) plays an important 
role in the high-energy regime of such machines. 
To account for the ISR effects, we also compute the corresponding 
ISR corrections to the considered processes. 
The corrected cross sections, including ISR effects for 
$\mu^+\mu^- \to H^\pm H^\mp V$, are given as follows:
\begin{eqnarray} 
\label{ISR}
\sigma^{\mu^+\mu^-\rightarrow 
H^{\pm}H^{\mp}V}_\text{ISR}(s)
&=& \int^1_0 dx \,H(x,s) \,
\sigma^{\mu^+\mu^-\rightarrow 
H^{\pm}H^{\mp}V}_\text{Tree}
\left( s(1-x) \right),
\end{eqnarray}
where $s$ stands for the center-of-mass
energy and $x$ is for the energy fraction 
of an emitted photon from muon. The radiator
function is taken in~\cite{Fujimoto:1990tb}
\begin{eqnarray} 
\label{HHSF}
H(x,s)&=&\Delta_2\beta x^{\beta-1}
-\Delta_1\beta\left(1-\frac{x}{2}\right)
\nonumber\\
&&+\frac{\beta^2}{8}\left[
-4(2-x)\log{x}-\frac{1+3(1-x)^2}{x}\log{(1-x)}-2x
\right].
\end{eqnarray}
In the formulas, we have 
\begin{eqnarray}
\label{ISRLoop}
\beta&=&\frac{2\alpha}{\pi}
\left(\ln \frac{s}{m_{\mu}^2} -1\right),
\quad ~~
\Delta_1= 1+\delta_1,~~
\Delta_2 = 1+\delta_1+\delta_2,\notag\\
\delta_1 &=& 
\frac{\alpha}{\pi}
\left(\frac{3}{2} \ln \frac{s}{m_{\mu}^2} 
+\frac{\pi^2}{3}-2\right),~~~~
\delta_2 = 
\left(\frac{\alpha }{\pi}
\ln \frac{s}{m_{\mu}^2}\right)^2
\left(-\frac{1}{18}\ln \frac{s}{m_{\mu}^2} 
+\frac{119}{72}-\frac{\pi^2}{3}
\right).\notag
\end{eqnarray}
Following the perturbative calculations of 
initial-state photon emission 
diagrams up to two-loop order~\cite{Fujimoto:1990tb}, 
in which the collinear approximation and the collinear
logarithmic enhancement expansions are applied, we note 
that the terms proportional to $\alpha^2$ in 
Eq.~(\ref{ISRLoop}) correspond to the two-loop 
contributions.

\begin{figure}[H]
\centering
\begin{tabular}{cc}
\includegraphics[width=8.5cm, height=9cm]
{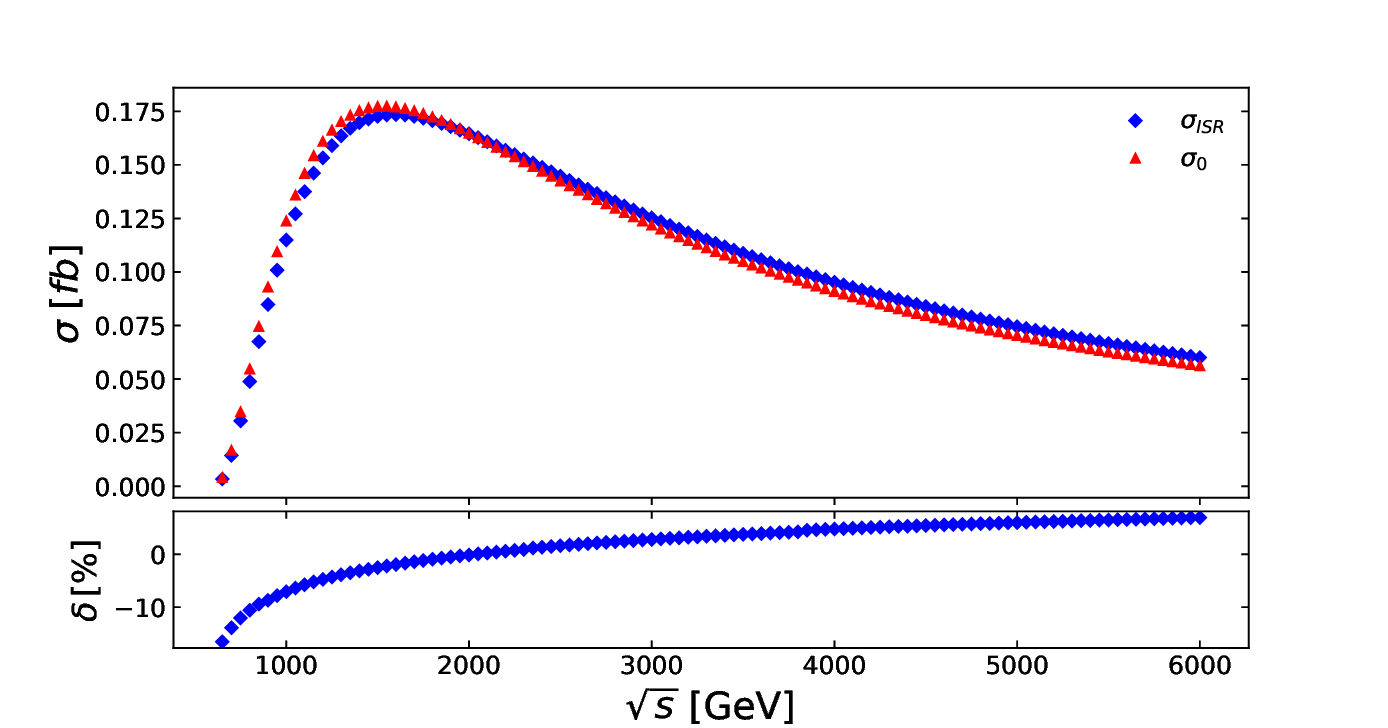}
&
\includegraphics[width=8.5cm, height=9cm]
{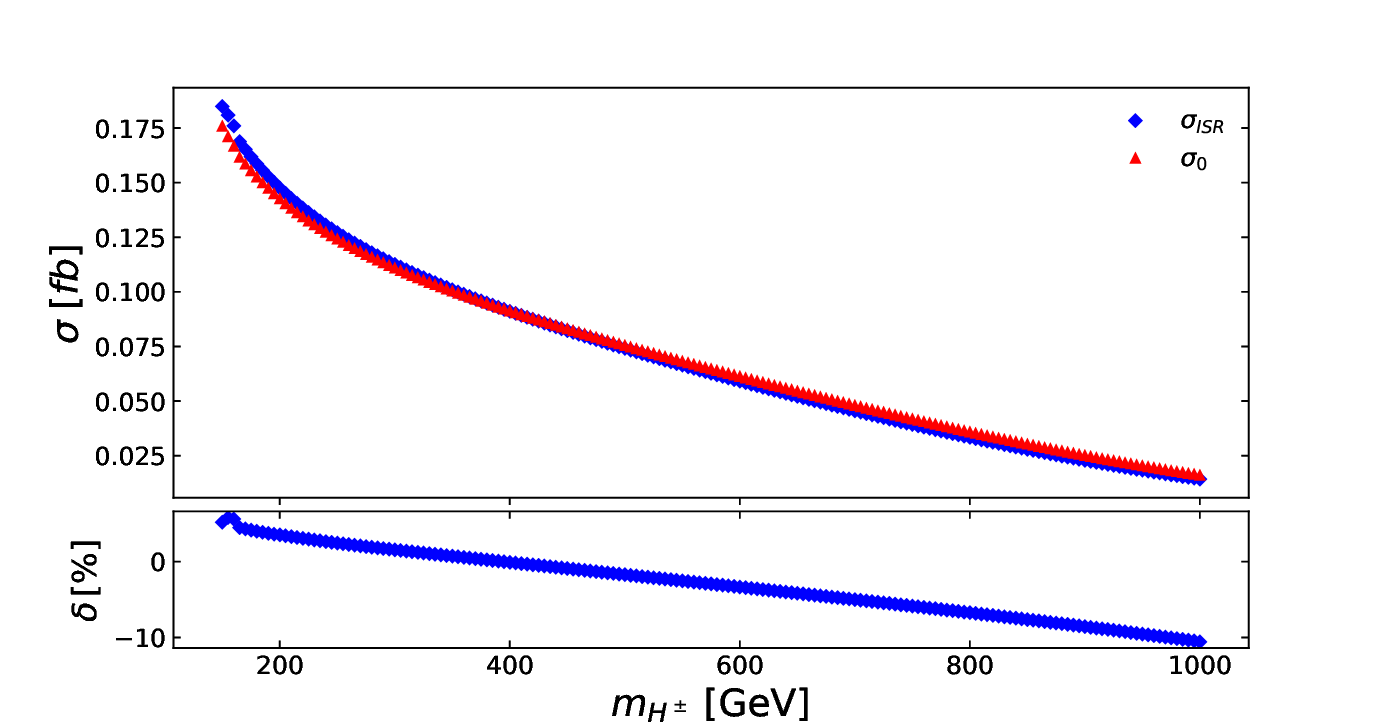}
\end{tabular}
\caption{\label{isreffect}
Effects of ISR on the production cross 
sections for $\mu^+\mu^- \to  
H^{\pm}H^{\mp}Z$ as fucntions of 
$\sqrt{s}$ (left) and fix $\sqrt{s}=3$ 
TeV with varying $m_{H^\pm}$ (right).}
\end{figure}

We first study the effects of ISR corrections
on the production cross sections. For this examination, 
we use the following benchmark point BP1 given in Table~\ref{BMpoints}, as a typical example. In left 
panel of Fig.~\ref{isreffect}, cross sections 
are shown with respect to the center-of-mass energy from
$\sqrt{s}=650~\text{GeV}$ to $\sqrt{s}=6~\text{TeV}$. 
While cross section at $\sqrt{s}=3000~\text{GeV}$ with
varying charged Higgs masses from $180$ GeV to $1000$ GeV
are studied in the right plot. 
In these Figures, the red line represents the tree-level 
cross sections without ISR corrections, while the blue 
line corresponds to the tree-level cross sections 
including ISR corrections. The corrections range up
to approximately $-10\%$ contributions. 

We now discuss the electroweak corrections to both
the signal processes and the corresponding Standard 
Model backgrounds. Based on the approximate Sudakov 
logarithmic behavior, the electroweak corrections 
can be estimated in the high-energy regime as 
follows~\cite{Kuhn:1999nn}:
\begin{eqnarray}
 \delta_{\mathrm{EW}} \approx 
-\,\frac{\alpha}{4\pi\,\sin^2\theta_W}\,
\kappa  \,\ln\!\left(\frac{s}{M_W^2}\right).
\end{eqnarray}
The effective coefficient is taken as 
$\kappa = 2,~3,~\cdots$ 
for scattering processes of the type 
$2 \to 2$, $2 \to 3$, and so on. 
By taking the following
input parameters, one can estimate 
electroweak corrections about $10\%$ 
contributions at $\sqrt{s}=3$ TeV.

For the SM background, we use 
MadGraph$5$MC@NLO~\cite{Alwall:2014hca, 
Frederix:2018nkq} for genarating the process 
$\mu^- \mu^+ \to t\bar{t} b\bar{b}
V,~WWhhV$ with $V=\gamma, Z$. In order to reduce
the SM background, we apply the following cuts:
\begin{eqnarray}
 p_T(b) \geq 20 \textrm{GeV}, \quad |\eta_b|\leq 2.4
 \quad p_T(\bar{b}) \geq 20 \textrm{GeV},  
 \quad |\eta_{\bar{b}}|\leq 2.4.
\end{eqnarray}
Furthermore, one applies $E_{\gamma}\geq 5$ GeV
and $|\cos\theta_{\gamma}|<0.98$ for 
external photon.

In this work, the significances are evaluated as
follows:
\begin{eqnarray}
 \mathcal{S} = \frac{N_{\textrm{S} }}
 {
 \sqrt{
 N_{\textrm{S} }
 + \varepsilon_B N_{\textrm{B} }
 }
 }.
\end{eqnarray}
Where $N_{\textrm{S/B}} = \mathcal{L} \sigma_{\textrm{S/B}}$ 
is corresponding for number of event and background. 
Where $\varepsilon_B$ explains for systematic uncertainty 
fraction on the background yield. 
We take $\varepsilon_B=1, 1.5$
in this work. 

\begin{figure}[H]
\centering
\begin{tabular}{cc}
$\mu^+\mu^- \to \gamma\gamma \to 
H^{\pm}H^{\mp}\gamma \to tbtb\gamma$
&
$\mu^+\mu^- \to \gamma\gamma \to 
H^{\pm}H^{\mp}\gamma \to tbtb \gamma$
\\
\includegraphics[width=8.5cm, height=7cm]
{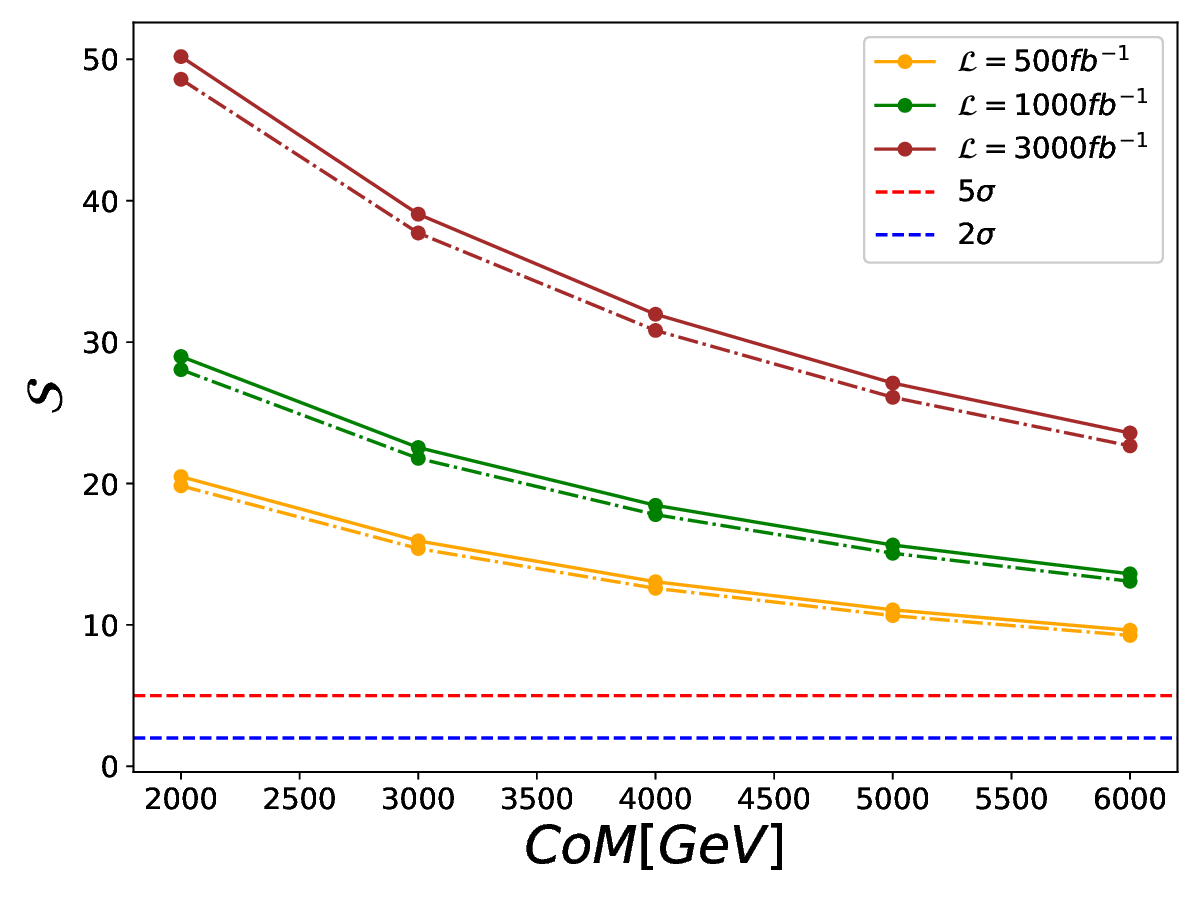}
&
\includegraphics[width=8.5cm, height=7cm]
{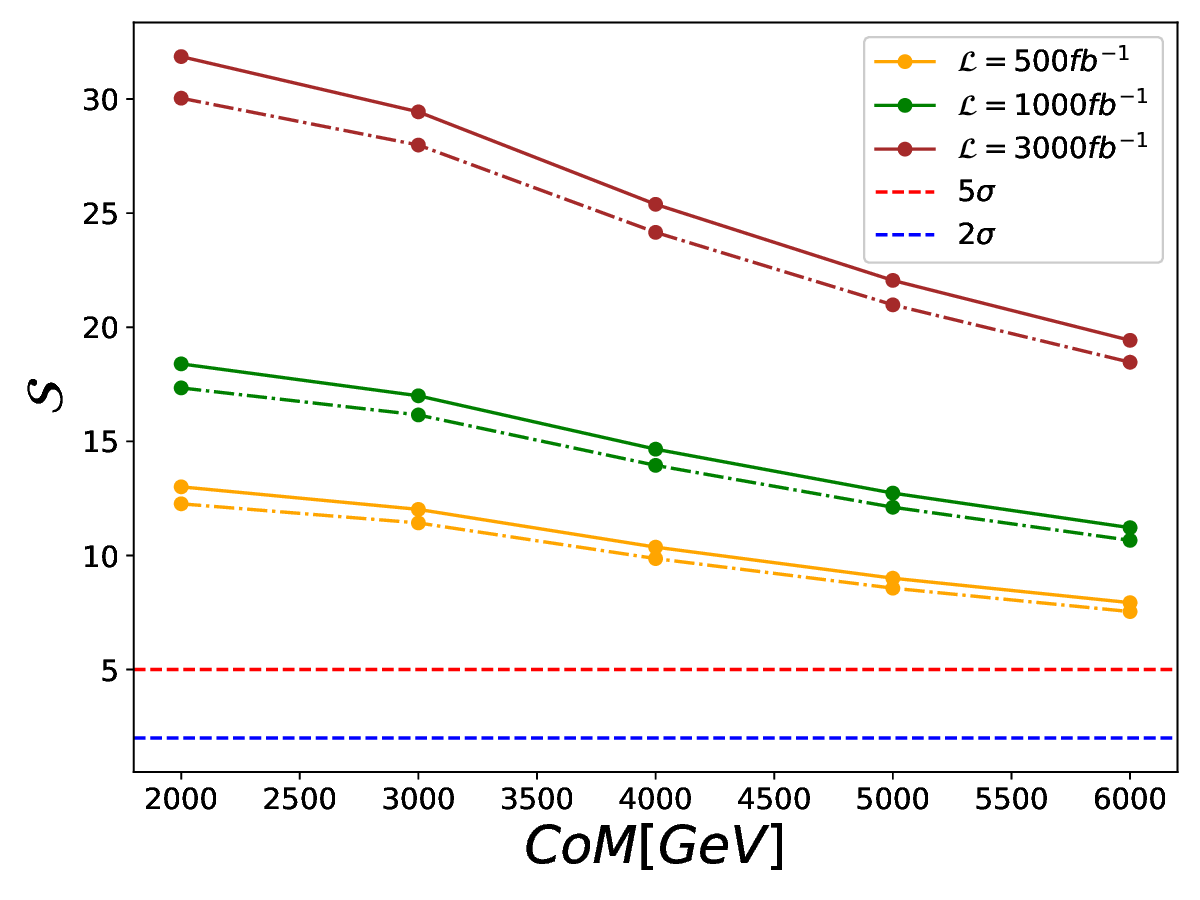}
\end{tabular}
\caption{\label{Signtbtbgam}
Signal significance for 
$\mu^+\mu^- \to \gamma\gamma \to 
H^{\pm}H^{\mp}\gamma \to tbtb\gamma$ 
at the benchmark points BP1 (left panel) 
and BP2 (right panel). 
}
\end{figure}

\begin{figure}[H]
\centering
\begin{tabular}{cc}
\hspace{0.5cm}
$\mu^+\mu^- \to \gamma\gamma \to 
H^{\pm}H^{\mp}\gamma \to WhWh\gamma$
&
\hspace{0.5cm}
$\mu^+\mu^- \to \gamma\gamma \to 
H^{\pm}H^{\mp}\gamma \to WhWh \gamma$
\\
\includegraphics[width=8.5cm, height=7cm]
{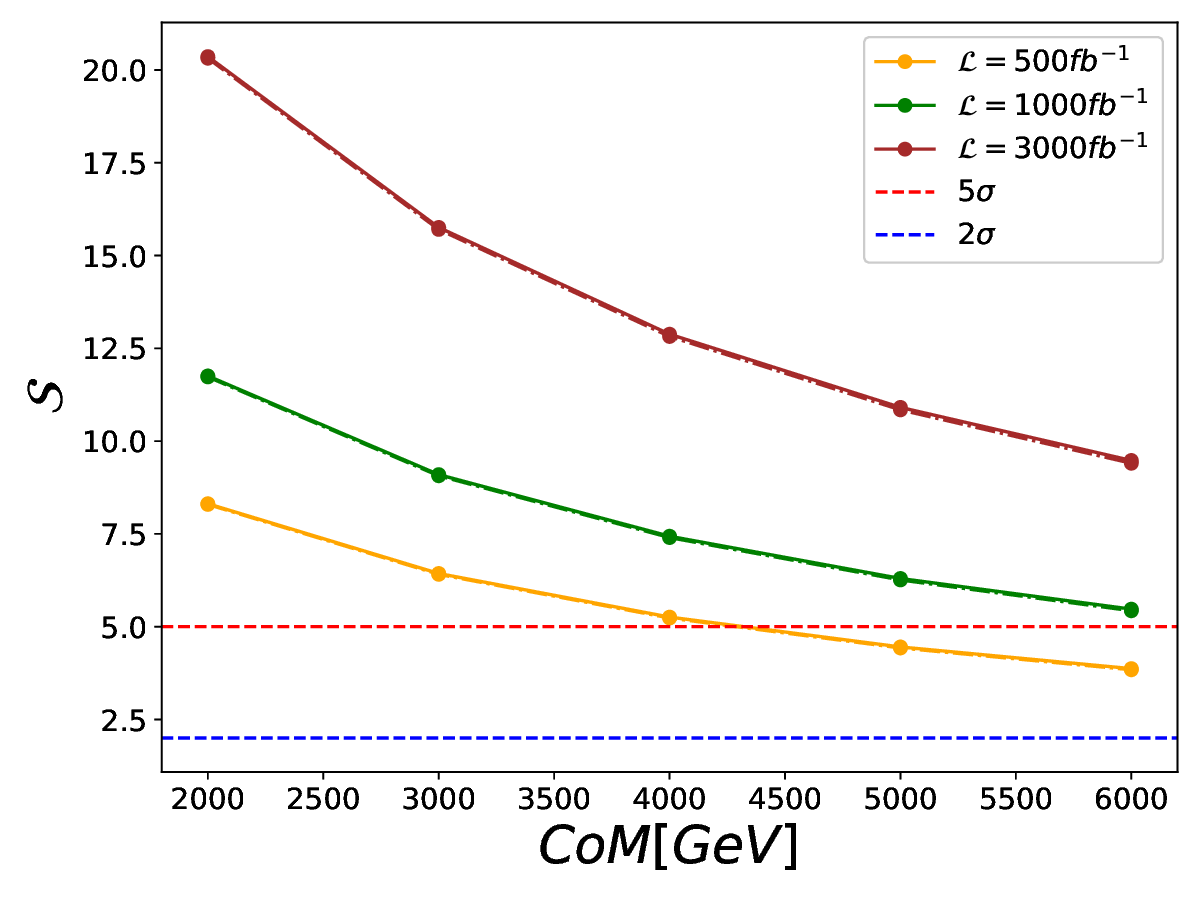}
&
\includegraphics[width=8.5cm, height=7cm]
{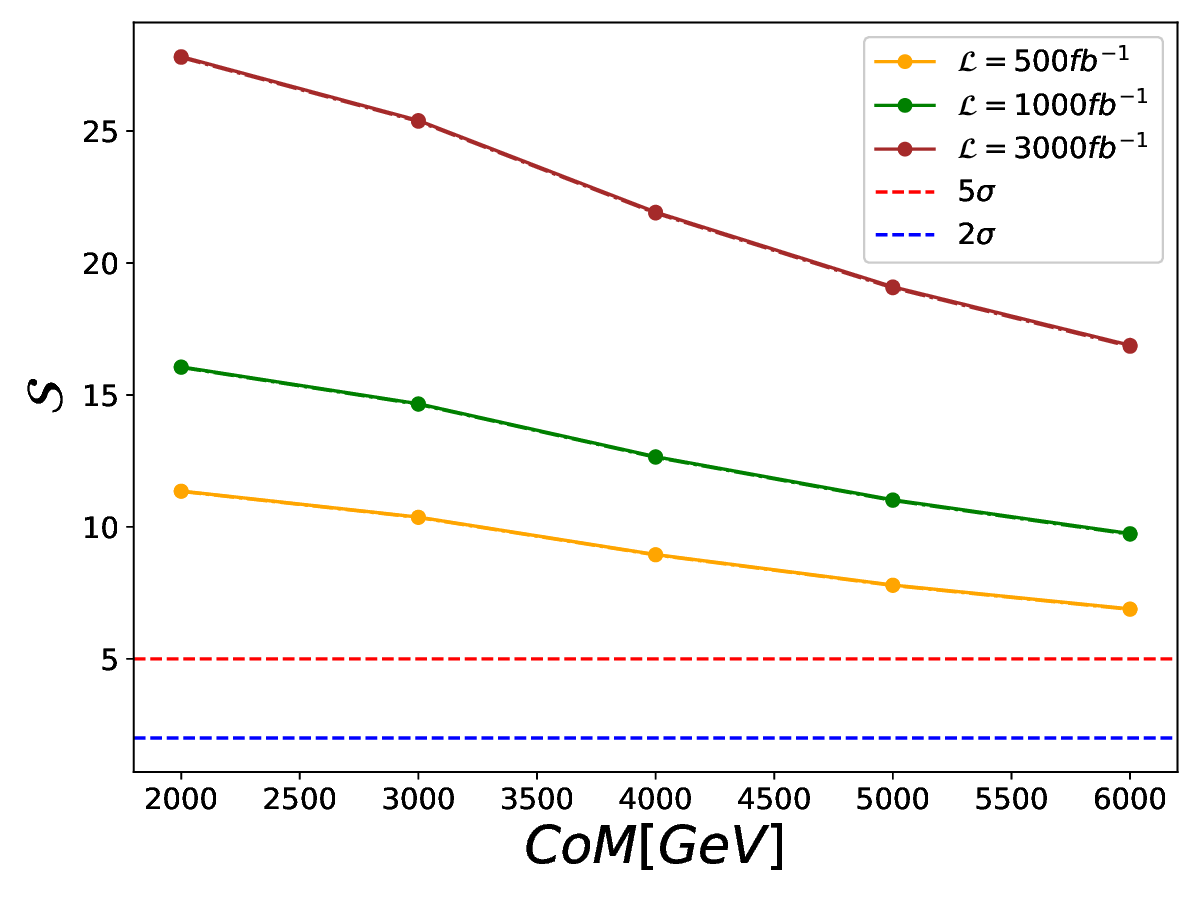}
\end{tabular}
\caption{\label{SignWhWhGam}
Signal significance for 
$\mu^+\mu^- \to \gamma\gamma \to 
H^{\pm}H^{\mp}\gamma \to WWhh\gamma$ 
at the benchmark points BP3 (left panel) 
and BP4 (right panel). 
}
\end{figure}

\begin{figure}[H]
\centering
\begin{tabular}{cc}
$\mu^+\mu^- \to 
H^{\pm}H^{\mp}Z \to tbtbZ$
&
$\mu^+\mu^- \to 
H^{\pm}H^{\mp}Z \to tbtb Z$
\\
\includegraphics[width=8.5cm, height=7cm]
{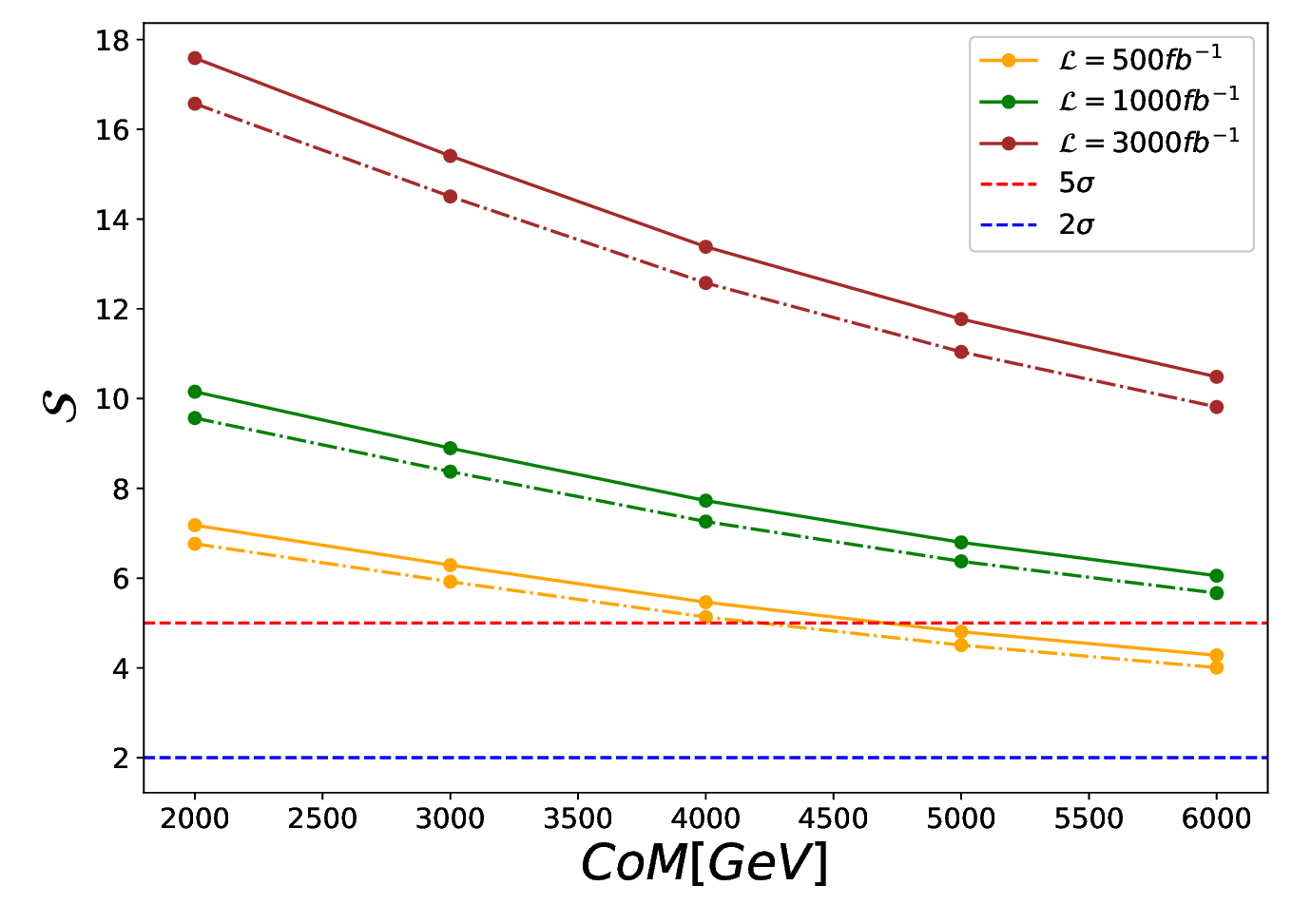}
&
\includegraphics[width=8.5cm, height=7cm]
{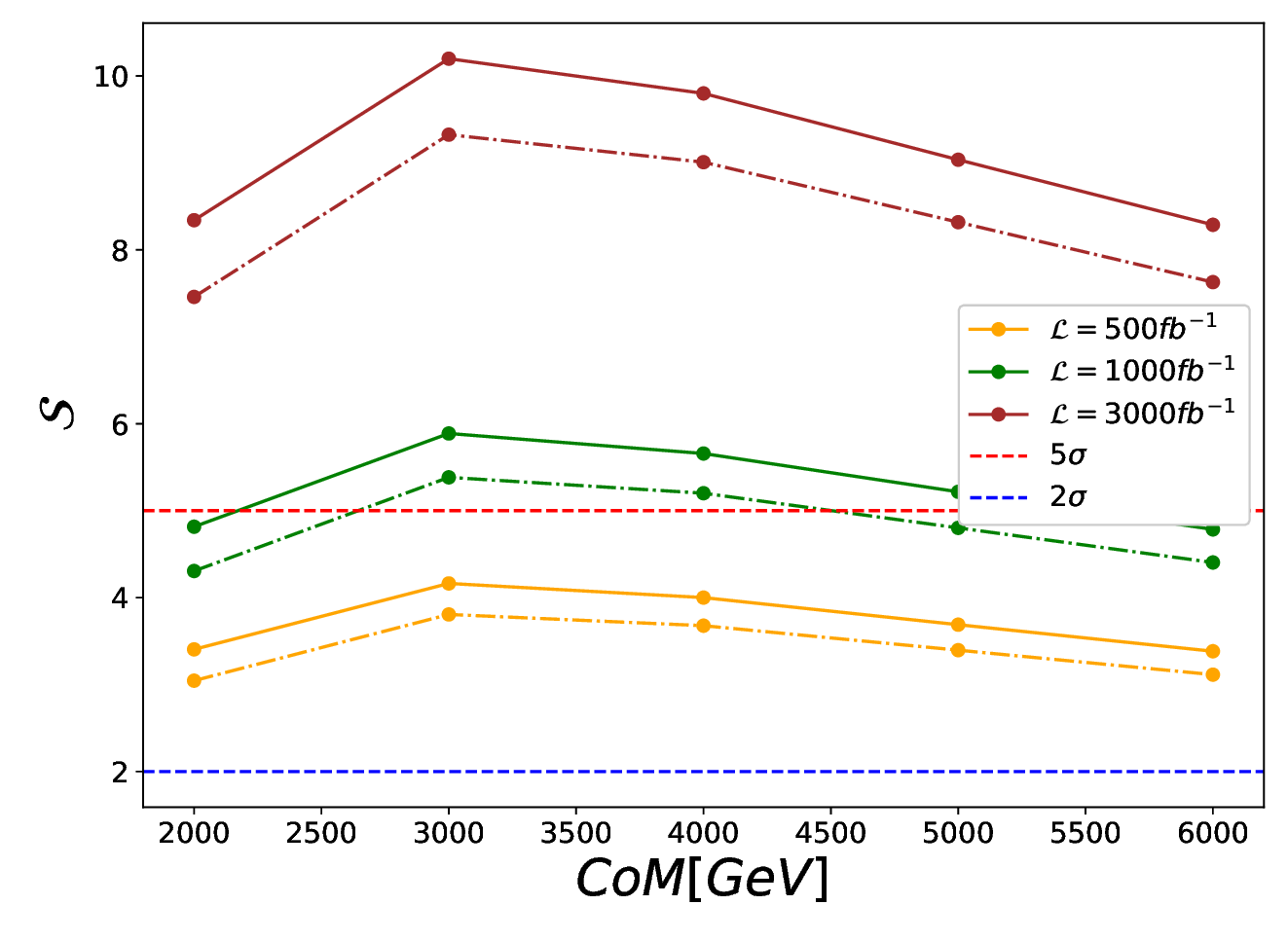}
\end{tabular}
\caption{\label{Smm2tbtbZ}
The signal significances of the process
$\mu^+\mu^- \to H^{\pm}H^{\mp}Z \to tbtbZ$
are studied for the benchmark points
BP1 (left) and BP2 (right). }
\end{figure}

In Fig.~\ref{Signtbtbgam}, we present the signal significance
for $\mu^+\mu^- \to \gamma\gamma \to H^{\pm}H^{\mp}\gamma
\to tbtb\gamma$ at the benchmark points BP1 (left panel)
and BP2 (right panel). The significances are computed by
varying the center-of-mass energy from 2~TeV to 6~TeV and
for different integrated luminosities of $\mathcal{L} = 500$
(yellow line), $\mathcal{L} =1000$ (green line) and
$\mathcal{L} =3000$~fb$^{-1}$ (red line). In the Figures,
the solid line corresponds to $\varepsilon_B = 1$, while
the dashed line represents $\varepsilon_B = 1.5$,
respectively. Our finding is that significances are above
$5\sigma$ for all ranges of center-of-mass energies.
Moreover, the signal significance reaches its maximum at
$\sqrt{s} = 2$~TeV and decreases as the center-of-mass
energy increases.
In Fig.~\ref{SignWhWhGam}, we examine 
the signal significance for 
$\mu^+\mu^- \to \gamma\gamma \to H^{\pm}H^{\mp}\gamma
\to WWhh\gamma$ at the benchmark points BP3 (left panel)
and BP4 (right panel). We use the same notations for
lines presented in these Figures as previous cases. 
In general, we observe
that the signal significances decrease with increasing
center-of-mass energy. For BP3, the significance exceeds
$5\sigma$ when $\sqrt{s} \leq 4~\text{TeV}$. Across all
energies considered, BP4 remains observable with a
significance above $5\sigma$.
The significances for the process $\mu^+\mu^- \to
H^{\pm}H^{\mp}Z \to tbtbZ$ are investigated at several
benchmark points indicated in the previous sections. Figure~\ref{Smm2tbtbZ} shows the
signal significances for this process at the benchmark point
BP1 (left) and BP2 (right). In the left panel, the results
show that the production can be observed with a significance
exceeding $5\sigma$ for integrated luminosities of
$\mathcal{L} \geq 1000~\text{fb}^{-1}$ for all energy ranges.
For $\mathcal{L} = 500~\text{fb}^{-1}$, the signal can be
tested with a significance above $5\sigma$ when
$\sqrt{s} \leq 4~\text{TeV}$. At other ranges of
center-of-mass energies, the production is difficult to test
at future muon–TeV colliders because its cross section is
smaller. In the right panel plot, the result indicates that
the production can be probed with a significance exceeding
$5\sigma$ for integrated luminosities of
$\mathcal{L}=3000~\text{fb}^{-1}$ for all energy ranges.
It is stressed that all notations for the lines shown in
these Figures are the same as in the previous cases.
Finally, we consider the significance for the production process
$\mu^+\mu^- \to H^{\pm}H^{\mp}Z \to WhWhZ$ at future muon–TeV colliders.
The signal significances are examined for two benchmark points, as shown
in the scatter plots presented in the previous Figures. In the left plot,
the significance for this process at BP3 is shown. It is found that the
signal can exceed $5\sigma$ for $\mathcal{L} = 3000~\text{fb}^{-1}$ in
the energy range from 2 to 4 TeV. In other cases, it is difficult to
test the production. In the right plot, the significance for this process
at BP4 is presented. The significance is enhanced in the low-energy region,
exceeding $5\sigma$ over the entire energy range for $\mathcal{L} \geq
1000~\text{fb}^{-1}$. Again, we use the same labels for the lines as in
previous plots.
\begin{figure}[H]
\centering
\begin{tabular}{cc}
$\mu^+\mu^- \to 
H^{\pm}H^{\mp}Z \to WWhhZ$
&
$\mu^+\mu^- \to
H^{\pm}H^{\mp}Z \to WWhhZ$
\\
\includegraphics[width=8.5cm, height=7cm]
{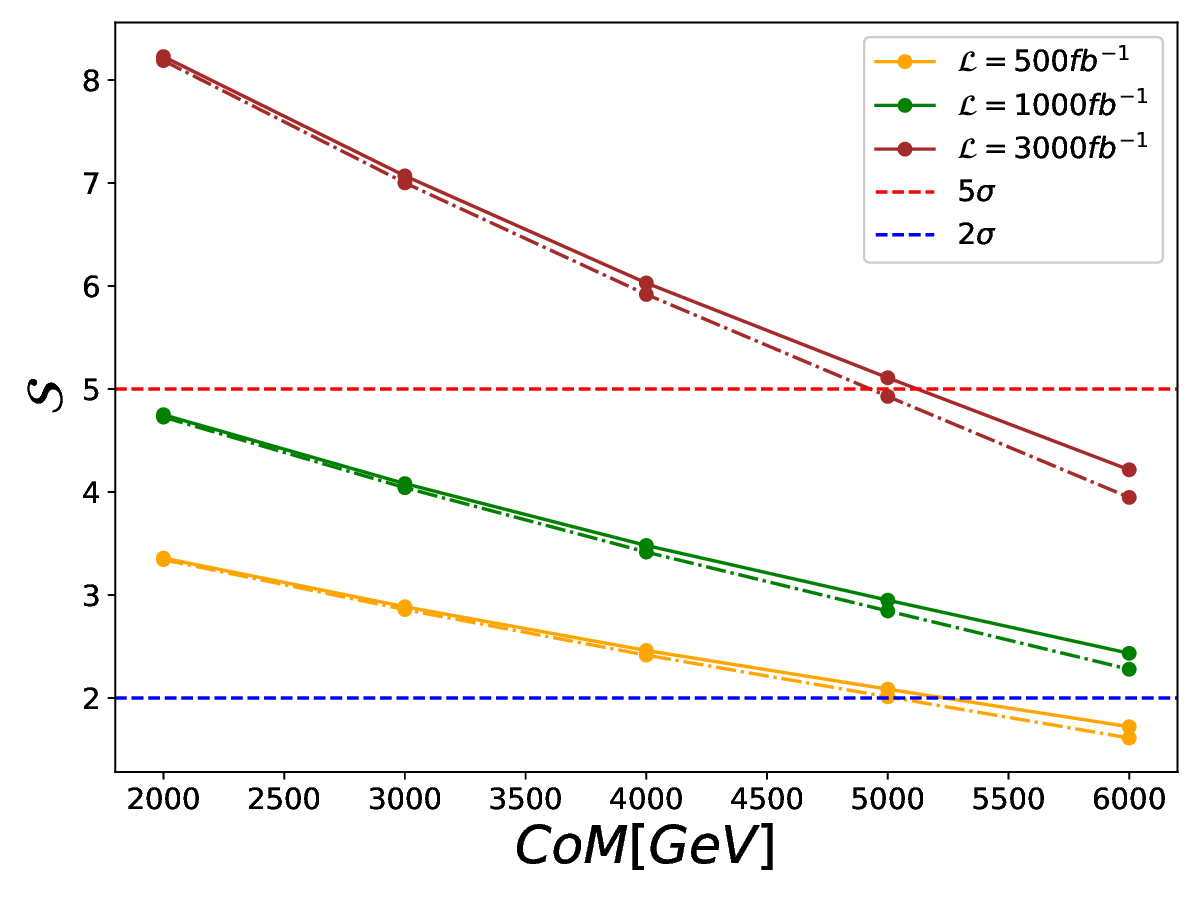}
&
\includegraphics[width=8.5cm, height=7cm]
{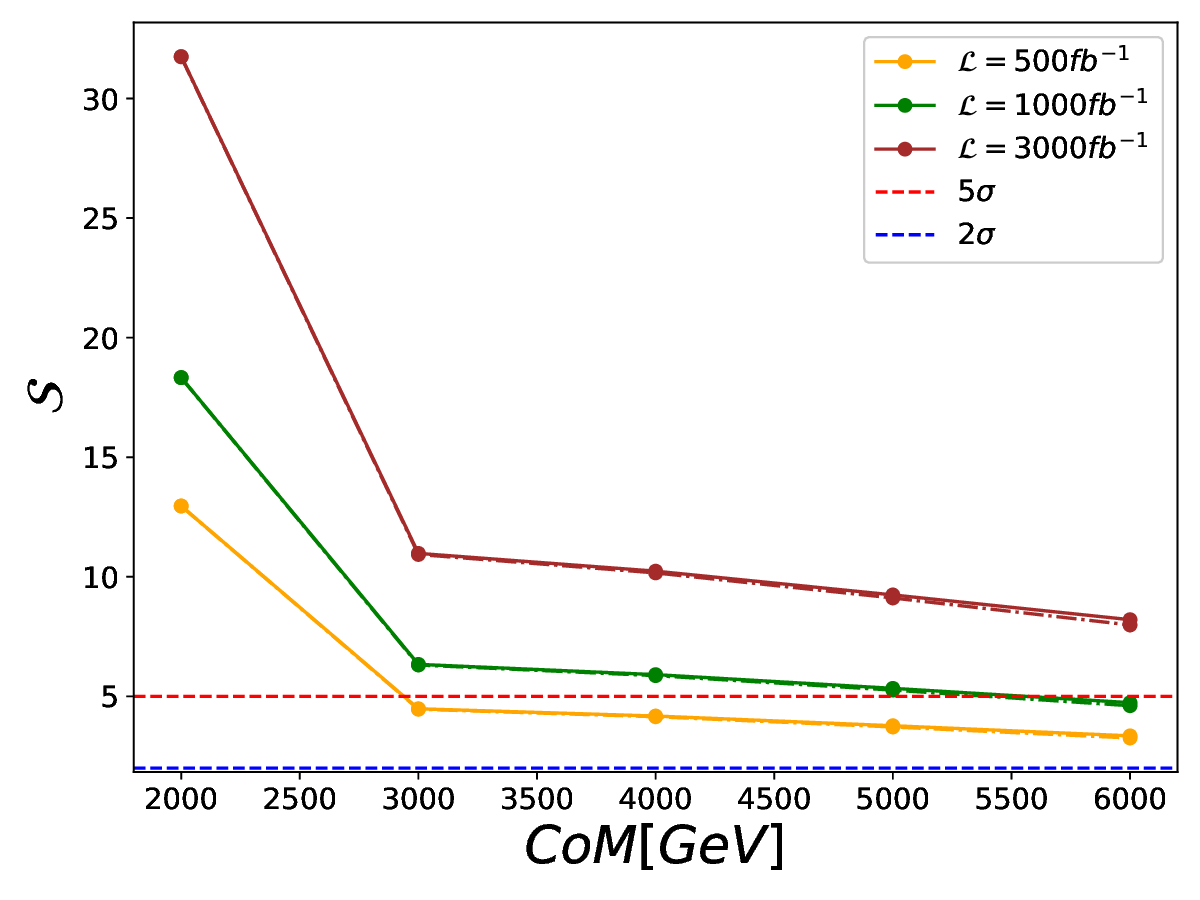}
\end{tabular}
\caption{\label{SmmWWhhZ}
The signal significances of the process
$\mu^+\mu^- \to H^{\pm}H^{\mp}Z \to WWhhZ$
are studied for the benchmark points
BP3 (left) and BP4 (right).
}
\end{figure}
\section{Conclusions} 
Charged Higgs pair production associated with 
neutral gauge bosons at
future multi--TeV muon colliders is first presented 
within the framework of THDM in this paper. 
In the phenomenological results for the
production processes, the parameter space of 
the Type-X THDM is updated in this work. From 
the viable parameter regions for the model 
under consideration, we calculate the production 
cross sections, including all two-body decay 
modes of the charged Higgs boson. It is worth 
emphasizing that all two-body decays of the charged 
Higgs include one-loop contributions, with two-loop 
QCD corrections also included for decays into two fermions.
The signal significances are also examined for several 
benchmark points within the updated parameter space.
It is worth stressing that the production cross 
sections used to compute the significances in this work 
include initial-state radiation corrections up to 
two-loop order. Thanks to the high integrated luminosity 
planned at future multi--TeV muon colliders, the signals 
are expected to be observable with a statistical 
significance of up to $5\sigma$.
\\

\noindent
{\bf Acknowledgment:}~
This research is funded by Vietnam
National Foundation for Science and
Technology Development (NAFOSTED) under
the grant number $103.01$-$2023.16$.
\subsection*{Process $\mu^+\mu^-
\to H^{\pm}H^{\mp} V$ for $V=\gamma,~Z$}
Tree-level Feynman diagrams are generated using {\tt FeynArts/FormCalc}. For the process $\mu^+\mu^- \to H^{\pm}H^{\mp} \gamma$, the external $Z$ boson is replaced by an external photon. Subsequently, diagrams $7$ and $8$ are omitted. For generality, 
we show the dominant Feynman diagrams for $\mu^+\mu^- \to H^{\pm}H^{\mp} Z$. It is emphasized that all Feynman diagrams
are retained in our calculations. However, to present the amplitude expressions in a compact form, we only display the amplitudes in the limit $m_{\mu} \rightarrow 0$ in the formulas below.
\begin{figure}[H]
\centering
\includegraphics[width=14cm, height=10cm]
{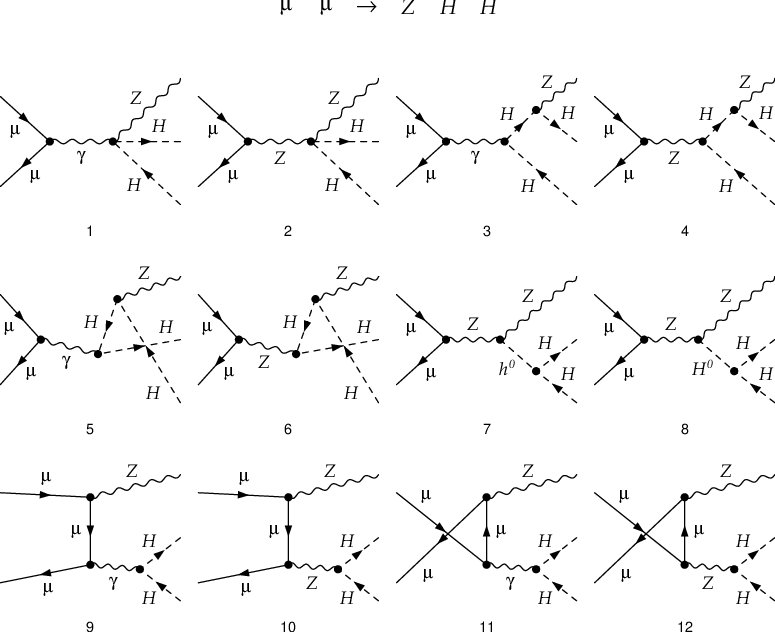}
\caption{\label{feymm2SSZ}
Tree-level Feynman diagrams are generated by
{\tt FeynArts/FormCalc}. For the process 
$\mu^+\mu^-
\to H^{\pm}H^{\mp} \gamma$, we replace the 
external $Z$ boson to external photon. 
Subsequently, 
we eliminate diagrams $7$ and $8$. }
\end{figure}
Amplitude for the process is presented as follows:
\begin{eqnarray}
i\mathcal{M} &=&\sum\limits_{k=1}^{8} 
F_k \cdot  \langle v| P_k |u \rangle.  
\end{eqnarray}
The tree-level form factor $F_k$ are functions 
of kinematic variables as:
$s=s_{12}= \left(k_1 + k_2\right)^2$, 
$s_{34} = \left(k_3 + k_4\right)^2$,
$t=t_{13} = \left(k_1 - k_3\right)^2$, 
$t_{14} = \left(k_1 - k_4\right)^2$, 
$t_{24} = \left(k_2 - k_4\right)^2$, 
$u=u_{23}= \left(k_2 - k_3\right)^2$.
The detail expressions for 
form factor $F_k$ is given by:
\begin{eqnarray}
F_1 &=& 
\dfrac{e g_{H^\pm H^\mp \gamma V}}
{s_{12}}
-
\dfrac{g^L_{{V f f}} 
g_{H^\pm H^\mp V}}
{s_{12} - m_Z^2}
\nonumber\\
&& 
-
\dfrac{(g_{{V f f}}^L)^2 \; 
g_{H^\pm H^\mp V} }
{\left(t_{13} - m_\mu^2\right)}
\dfrac{
\left(m_{H^\pm}^2 + m_Z^2 
- s_{34} - t_{13}\right)}
{
\left(3 m_{H^\pm}^2 
+ 2 m_\mu^2 - s_{34} 
- t_{14} - t_{24} \right)
}
\nonumber\\
&& 
+
\dfrac{e^2 \;g_{{Vff}}^L 
}{\left(t_{13} - m_\mu^2\right)}
\dfrac{
\left(m_{H^\pm}^2 + m_Z^2 
- s_{34} - t_{13} \right)
}
{\left(3 m_{H^\pm}^2 + 2 m_\mu^2 
+ m_Z^2 - s_{34} 
- t_{14} - t_{24}\right)}
\nonumber\\
&& +
\dfrac{g^L_{{V f f}}
}{\left(s_{12} - m_Z^2\right)}
\sum\limits_{\phi_k=h,H}
\dfrac{
g_{\phi_k H^\pm H^\mp}\; g_{\phi_k V V} 
}
{ 
\left(m_{\phi_k}^2 - 3 m_{H^\pm}^2 
- 2 m_\mu - m_Z^2 
+ s_{34} + t_{14} + t_{24}\right)
}
\nonumber\\
&& 
+
\dfrac{(g_{{Vff}}^L)^2 
g_{H^\pm H^\mp V} 
}
{\left(u_{23} - m_\mu^2\right)
}
\dfrac{
\left(m_{H^\pm}^2 + m_Z^2 
- s_{34} - u_{23}\right)}
{
\left(3 m_{H^\pm}^2 + 2 m_\mu^2 
- s_{34} - t_{14} - t_{24}\right)
}
\nonumber\\
&& 
-
\frac{e^2\; g^L_{{Vff}} 
}{\left(u_{23} - m_\mu^2\right)}
\dfrac{
\left(m_{H^\pm}^2 + m_Z^2 
- s_{34} - u_{23}\right)
}
{
\left(3 m_{H^\pm}^2 + 2 m_\mu^2 
+ m_Z^2 - s_{34} - t_{14} 
- t_{24}\right)},
\end{eqnarray}
\begin{eqnarray}
F_2 &=& 
\dfrac{e g_{H^\pm H^\mp \gamma V}}
{s_{12} }
-
\dfrac{(g^R_{{V f f}})^2 
}
{\left(t_{13} - m_\mu^2\right)
}
\dfrac{
g_{H^\pm H^\mp V}
\left(m_{H^\pm}^2 
+ m_Z^2 - s_{34} - t_{13} \right)
}
{
\left(3 m_{H^\pm}^2 
+ 2 m_\mu^2 - s_{34} 
- t_{14} - t_{24} \right)}
\nonumber\\
&& -
\dfrac{g^R_{{V f f}} 
g_{H^\pm H^\mp V}}
{s_{12} - m_Z^2}
+
\dfrac{e^2\; g^R_{{Vff}} 
}
{\left(t_{13} - m_\mu^2\right)
}
\dfrac{
\left(m_{H^\pm}^2 + m_Z^2 
- s_{34} - t_{13} \right)
}
{
\left(3 m_{H^\pm}^2 + 2 m_\mu^2 
+ m_Z^2 - s_{34} 
- t_{14} - t_{24}\right)}
\nonumber\\
&& 
+
\dfrac{g^R_{{Vff}} 
}
{
\left(s_{12} - m_Z^2\right)
}
\sum\limits_{\phi_k=h,H}
\dfrac{
g_{\phi_k H^\pm H^\mp} 
g_{\phi_kVV}}
{
\left(m_{\phi_k}^2 - 3 m_{H^\pm}^2 - 2 m_\mu 
- m_Z^2 + s_{34} 
+ t_{14} + t_{24}\right)}
\nonumber\\
&& 
+
\dfrac{(g^R_{{Vff}})^2
g_{H^\pm H^\mp V} 
}
{\left(u_{23} - m_\mu^2\right)}
\dfrac{
\left(m_{H^\pm}^2 + m_Z^2 
- s_{34} - u_{23}\right)}
{
\left(3 m_{H^\pm}^2
+ 2 m_\mu^2 - s_{34} 
- t_{14} - t_{24}\right)}
\nonumber\\
&& 
-
\dfrac{e^2\; g^R_{{Vff}} 
}
{\left(u_{23} - m_\mu^2\right)
}
\dfrac{
\left(m_{H^\pm}^2 + m_Z^2 
- s_{34} - u_{23}\right)}
{
\left(3 m_{H^\pm}^2 + 2 m_\mu^2 
+ m_Z^2 - s_{34} 
- t_{14} - t_{24}\right)},
\nonumber\\
F_3 &=&
\dfrac{4\; g_{H^\pm H^\mp V}^2 \; 
g^R_{{V f f}}
\left(
\varepsilon^*(k_3) \cdot k_4
\right)}
{\left(s_{34} - m_{H^\pm}^2\right)
\left(s_{12} - m_Z^2\right)}
-
\dfrac{4e^2\; g_{H^\pm H^\mp V} 
\left(\varepsilon^*(k_3) \cdot k_4\right)}
{\left(s_{34} - m_{H^\pm}^2\right)s_{12}}
\nonumber\\
&& - 
\frac{4 (g^R_{{Vff}})^2 
g_{H^\pm H^\mp V} 
\left(\varepsilon^*(k_3) \cdot k_1\right)}
{\left(t - m_\mu^2\right)\left(3 m_{H^\pm}^2 
+ 2 m_\mu^2 - s_{34} - t_{24} - t_{14}\right)}
\nonumber\\
&& +
\frac{4e^2\; g^R_{{V f f}}\; 
\left(\varepsilon^*(k_3) \cdot k_1\right)}
{\left(t - m_\mu^2\right)
\left(3 m_{H^\pm}^2 + 2 m_\mu^2 
+ m_Z^2 - s_{34} - t_{24} - t_{14}\right)}
\nonumber\\
&& -
\frac{4 (g^R_{{V ff }})^2 g_{H^\pm H^\mp V} 
\left(\varepsilon^*(k_3) \cdot k_2\right)}
{\left(u - m_\mu^2\right)\left(3 m_{H^\pm}^2
+ 2 m_\mu^2 - s_{34} - t_{24} - t_{14}\right)}
\nonumber\\
&& + 
\frac{4e^2\;  g^R_{{V f f}} 
\left(\varepsilon^*(k_3) \cdot k_2\right)}
{\left(u - m_\mu^2\right)
\left(3 m_{H^\pm}^2 + 2 m_\mu^2 
+ m_Z^2 - s_{34} - t_{24} 
- t_{14}\right)}
\nonumber\\
&& +
\frac{4 g_{H^\pm H^\mp V}
\left(\varepsilon^*(k_3) \cdot k_5\right)
\left[e^2\; \left(s - m_Z^2\right)
- g^R_{{V f f}} g_{H^\pm H^\mp V}\; s\right]}
{s\left(s - m_Z^2\right)\left(m_{H^\pm}^2
+ 2 m_\mu^2 + 2 m_Z^2 - s_{34} - t - u\right)},
\nonumber\\
F_4 &=&
\frac{4 g_{H^\pm H^\mp V}^2 g^L_{{V f f}} 
\left(\varepsilon^*(k_3) \cdot k_4\right)}
{\left(s_{34} - m_{H^\pm}^2\right)
\left(s - m_Z^2\right)}
-
\frac{4e^2 g_{H^\pm H^\mp V} \; 
\left(\varepsilon^*(k_3) \cdot k_4\right)}
{\left(s_{34} - m_{H^\pm}^2\right)s}
\nonumber\\
&& -
\frac{4 (g^L_{{V f f}})^2 g_{H^\pm H^\mp V}
\left(\varepsilon^*(k_3) \cdot k_1\right)}
{\left(t - m_\mu^2\right)
\left(3 m_{H^\pm}^2 + 2 m_\mu^2 - s_{34} 
- t_{24} - t_{14}\right)}
\nonumber\\
&& +
\frac{4e^2\;  g^L_{{V f f}} \; 
\left(\varepsilon^*(k_3) \cdot k_1\right)}
{\left(t - m_\mu^2\right)
\left(3 m_{H^\pm}^2 + 2 m_\mu^2 
+ m_Z^2 - s_{34} - t_{24} - t_{14}\right)}
\nonumber\\
&& -
\frac{4 (g^L_{{V ff }})^2 
g_{H^\pm H^\mp V} 
\left(\varepsilon^*(k_3) \cdot k_2\right)}
{\left(u - m_\mu^2\right)
\left(3 m_{H^\pm}^2 + 2 m_\mu^2 
- s_{34} - t_{24} - t_{14}\right)}
\nonumber\\
&& 
+ 
\frac{4e^2\; g^L_{{V f f}} \;
\left(\varepsilon^*(k_3) \cdot k_2\right)}
{\left(u - m_\mu^2\right)
\left(3 m_{H^\pm}^2 + 2 m_\mu^2 
+ m_Z^2 - s_{34} - t_{24} - t_{14}\right)}
\nonumber\\
&& 
+
\frac{4 g_{H^\pm H^\mp V}
\left(\varepsilon^*(k_3) \cdot k_5\right)
\left[e^2 \; \left(s - m_Z^2\right)
- g^L_{{V f f}} g_{H^\pm H^\mp V} \; s\right]}
{s\left(s - m_Z^2\right)
\left(m_{H^\pm}^2 + 2 m_\mu^2 
+ 2 m_Z^2 - s_{34} - t - u\right)},
\end{eqnarray}
\begin{eqnarray}
F_5 &=&
\frac{4 g_{H^\pm H^\mp V}^2 g^L_{{Vff}} 
\left(\varepsilon^*(k_3) \cdot k_4\right)}
{\left(s_{34} - m_{H^\pm}^2\right)
\left(s - m_Z^2\right)}
-
\frac{4e^2\; g_{H^\pm H^\mp V} 
\left(\varepsilon^*(k_3) \cdot k_4\right)}
{\left(s_{34} - m_{H^\pm}^2\right)s}
\nonumber\\
&& + 
\left(
\frac{\varepsilon^*(k_3) \cdot k_4}{u - m_\mu^2}
-
\frac{\varepsilon^*(k_3) \cdot k_4}{t - m_\mu^2}
\right)
\times
\nonumber\\
&&
\times 
\left[
\frac{2 (g^L_{{V f f}})^2 
g_{H^\pm H^\mp V}}
{3 m_{H^\pm}^2 + 2 m_\mu^2 
- s_{34} - t_{14} - t_{24}}
-
\frac{2e^2\; g^L_{{V f f}} }
{3 m_{H^\pm}^2 + 2 m_\mu^2 
+ m_Z^2 - s_{34} - t_{14} - t_{24}}
\right],
\nonumber\\
F_6 &=&
\frac{4 g_{H^\pm H^\mp V}^2 g^R_{{V f f}} 
\left(\varepsilon^*(k_3) \cdot k_4\right)}
{\left(s_{34} - m_{H^\pm}^2\right)
\left(s - m_Z^2\right)}
-
\frac{4e^2 g_{H^\pm H^\mp V} \;
\left(\varepsilon^*(k_3) \cdot k_4\right)}
{\left(s_{34} - m_{H^\pm}^2\right)s}
\nonumber\\
&& + 
\left(
\frac{\varepsilon^*(k_3) \cdot k_4}{u - m_\mu^2}
-
\frac{\varepsilon^*(k_3) \cdot k_4}{t - m_\mu^2}
\right)
\times 
\nonumber\\
&& \times
\left[
\frac{2 (g^R_{{Vff}})^2 g_{H^\pm H^\mp V}}
{3 m_{H^\pm}^2 + 2 m_\mu^2 
- s_{34} - t_{14} - t_{24}}
-
\frac{2e^2\; g^R_{{Vff}} }
{3 m_{H^\pm}^2 + 2 m_\mu^2 + m_Z^2 
- s_{34} - t_{14} - t_{24}}
\right],
\nonumber\\
F_7 &=&
\left(\frac{1}{u - m_\mu^2} 
+ \frac{1}{t - m_\mu^2}\right)
\times 
\nonumber\\
&& \times
\left[
\frac{2 (g^R_{{Vff}})^2 g_{H^\pm  H^\mp V}}
{3 m_{H^\pm}^2 + 2 m_\mu^2 - s_{34} 
- t_{14} - t_{24}}
-
\frac{2e^2 g^R_{{Vff}} }
{3 m_{H^\pm}^2 + 2 m_\mu^2 + m_Z^2 
- s_{34} - t_{14} - t_{24}} \right],
\nonumber\\
F_8 &=&
\left(
\frac{1}{u - m_\mu^2} 
+ \frac{1}{t - m_\mu^2}
\right)
\times 
\nonumber\\
&& \times
\left[
\frac{2 (g^L_{{Vff}})^2 g_{H^\pm H^\mp V}}
{3 m_{H^\pm}^2 + 2 m_\mu^2 - s_{34} 
- t_{14} - t_{24}}
-
\frac{2e^2\; g^L_{{Vff}} }
{3 m_{H^\pm}^2 + 2 m_\mu^2 + m_Z^2 
- s_{34} - t_{14} - t_{24}}
\right],
\end{eqnarray}
The operators $P_k$ are also taken 
the form of $P_1 = P_L \slashed{\varepsilon^*}(k_3)$, 
$P_2 = P_R \slashed{\varepsilon^*}(k_3)$,
$P_3 
= P_R \slashed{k_4}$, $P_4 = P_L \slashed{k_4}$, 
$P_5 = P_L \slashed{k_3}$, 
$P_6 = P_R \slashed{k_3}$, 
$P_7 = - P_R \slashed{\varepsilon^*}(k_3) 
\slashed{k_3} \slashed{k_4}$, 
$P_8 = - P_L \slashed{\varepsilon^*}(k_3)
\slashed{k_3} \slashed{k_4}$. 
Here, $\varepsilon^*(k_3)$ is polarization vector 
of external vector $V$. 
When we cosider $V=\gamma$, 
we take $g_{H^\pm H^\mp \gamma V}
=g_{H^\pm H^\mp \gamma\gamma}$, 
$g_{H^\pm H^\mp V}=g_{H^\pm H^\mp \gamma}$ 
and $g_{hZV}=g_{HZV}=0$. One then arrives 
at the corresponding amplitude for the 
process $\mu^+\mu^- \to H^{\pm}H^{\mp} \gamma$. 
Cross section is calculated: 
\begin{eqnarray}
\sigma(s) = \frac{1}{\textrm{flux}}
\int d\Phi_3 \frac{1}{4}\sum\limits_{\textrm{spin}}
|\mathcal{M}_{\mu^-\mu^+\to H^{\pm}H^{\mp} V}|^2.
\end{eqnarray}
\subsection*{Partonic process $\gamma\gamma
\to H^{\pm}H^{\mp} V$ for $V=\gamma,~Z$}
\begin{figure}[H]
\centering
\includegraphics[width=14cm, height=10cm]
{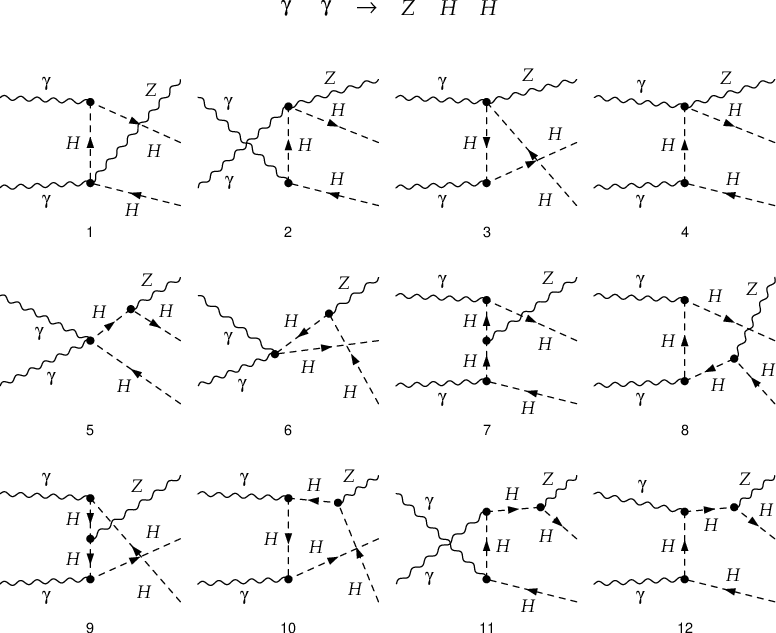}
\caption{\label{feygg2SSZ}
Tree-level Feynman diagrams are generated by {\tt FeynArts/FormCalc}.
The partonic process $\gamma\gamma \to H^{\pm}H^{\mp}\gamma$ 
will be obtained by approximately replacing the external
$Z$ by a photon.
}
\end{figure}

Tree-level Feynman diagrams for the process
$\gamma\gamma \to H^{\pm}H^{\mp} V$ with 
$V=\gamma,~Z$ are presented in Fig.~\ref{feygg2SSZ}. 
It is noted that we take $V=Z$ for generality.
Tree-level amplitude for partonic process 
$\gamma\gamma \to H^{\pm}H^{\mp} V$ 
is expressed as follows:
\begin{eqnarray}
i\mathcal{M}
&=&
\dfrac{
2
\left( 
\varepsilon(k_1) \cdot k_5
\right)
}
{m_{H^\pm}^2 + m_Z^2 - s_{12} - t_{13} - t_{14}}
\times 
\\
&& \times 
\Big[
e\; g_{H^\pm H^\mp V\gamma} 
\cdot \left(
\varepsilon(k_2) \cdot k_5
\right)
-
c_1
\dfrac{(4 e^2) \;
g_{H^\pm H^\mp V}}
{s_{34} - m_{H^\pm}^2} 
\cdot 
\left( \varepsilon^*(k_3) \cdot k_4  
\right)
\Big]
\nonumber\\
&&
- (2e)
g_{H^\pm H^\mp V \gamma}
\Big[
\dfrac{
\left(
\varepsilon(k_1) 
\cdot \varepsilon^*(k_3)
\right)
\left(
\varepsilon(k_2) \cdot k_5
\right)
}{t_{14} - m_{H^\pm}^2}
+
\dfrac{
\left( 
\varepsilon(k_1) 
\cdot k_4
\right)
\left(
\varepsilon(k_2) 
\cdot \varepsilon^*(k_3)
\right)
}{t_{24} - m_{H^\pm}^2}
\Big]
\nonumber\\
&&
+
\dfrac{
2
\left(
\varepsilon(k_2) \cdot k_4
\right)
}{m_{H^\pm}^2 + m_Z^2 - 
s_{12} - t_{24} - u_{23}}
\times 
\nonumber
\\
&&
\times 
\Big[
e\; g_{H^\pm H^\mp V \gamma} 
\left(
\varepsilon(k_1) \cdot k_4
\right)
-
(4e^2)
g_{H^\pm H^\mp V} 
\Big(
c_2
\frac{
\varepsilon^*(k_3) \cdot k_4
}{s_{34} - m_{H^\pm}^2}
+c_6
\frac{\varepsilon(k_1) \cdot \varepsilon^*(k_3)}
{t_{14} - m_{H^\pm}^2}
\Big) 
\Big]
\nonumber\\
&& +(4e^2)\; 
g_{H^\pm H^\mp V} 
\times 
\nonumber\\
&&
\times 
\left[
\frac{ 
\left(
\varepsilon(k_1) \cdot \varepsilon(k_2)
\right)
\left(
\varepsilon^*(k_3) \cdot k_4
\right)
}{m_{H^\pm}^2 - s_{34}}
+ c_4 
\frac{ 2
\left(
\varepsilon(k_2) \cdot \varepsilon^*(k_3)
\right)
\left(
\varepsilon(k_1) \cdot k_5
\right)
}
{\left(m_{H^\pm}^2 + m_Z^2 
- s_{12} - t_{13} - t_{14}\right)
\left(m_{H^\pm}^2 - t_{24}\right)}
\right.\nonumber\\
&&\left.
-
\frac{2
\left(
\varepsilon^*(k_3) \cdot k_5
\right)
}
{s_{34} + t_{13} + u_{23} 
- m_{H^\pm}^2 - 2 m_Z^2}
\right.
\times 
\nonumber
\\
&&
\hspace{4cm}
\left.
\times 
\left(
\frac{
\varepsilon(k_1) 
\cdot \varepsilon(k_2)
}{2}
- c_3
\frac{
\varepsilon(k_1) 
\cdot \varepsilon^*(k_3)
}
{m_{H^\pm}^2 - t_{14}}
-
c_5
\frac{
\varepsilon(k_2) 
\cdot \varepsilon^*(k_3)
}
{m_{H^\pm}^2 - t_{24}}
\right)
\right].
\nonumber
\end{eqnarray}
The polarization vectors of 
incoming photons are $\varepsilon(k_1)$, 
$\varepsilon(k_2)$ and of external vector 
$V$ is $ \varepsilon^*(k_3)$.
In the amplitude, corresponding 
coefficients are given by
\begin{eqnarray}
c_1 &=& \varepsilon(k_2) \cdot k_1
 -\varepsilon(k_2) \cdot k_4,  \\
c_2 &=& \varepsilon(k_1) \cdot k_2
 -  \varepsilon(k_1) \cdot k_5, \\
c_3 &=& \varepsilon^*(k_3) \cdot k_1
 - \varepsilon^*(k_3) \cdot k_5, \\
c_4 &=&  \varepsilon(k_2) \cdot k_1
 - \varepsilon(k_2) \cdot k_5, \\
c_5 &=& \varepsilon^*(k_3) \cdot k_1
 - \varepsilon^*(k_3) \cdot k_4, \\
c_6 &=& \varepsilon(k_1) \cdot k_2
 - \varepsilon(k_1) \cdot k_4. 
\end{eqnarray}
In the case of $V = \gamma$, we set the approximately
couplings $g_{H^\pm H^\mp V}=g_{H^\pm H^\mp \gamma}=e$
and $g_{H^\pm H^\mp V \gamma} = g_{H^\pm H^\mp \gamma \gamma} 
=2e^2$. 
Cross section is calculated by 
\begin{eqnarray}
 \sigma(s) = \frac{1}{\textrm{flux}}
 \int d\Phi_3 \frac{1}{4}\sum\limits_{\textrm{pol.}}
 |\mathcal{M}_{\gamma\gamma \to H^{\pm}H^{\mp} V }|^2.
\end{eqnarray}
\end{document}